\documentclass[apj]{emulateapj}
\usepackage{mathtools}
\usepackage{natbib}
\usepackage{hyperref}
\usepackage{graphicx,xspace,color,longtable}
\usepackage{rotate,amssymb,amsmath,shadow,bezier,curves}
\usepackage[flushleft]{threeparttable}

\shorttitle{GalWeight: An Effective Weighting Technique}
\shortauthors{Abdullah, Wilson \& Klypin 2017}

\begin{document}

\title{GalWeight: A New and Effective Weighting Technique for Determining Galaxy Cluster and Group Membership}

\author{Mohamed H. Abdullah}
\affil{Department of Physics and Astronomy, University of California Riverside, 900 University Avenue, Riverside, CA 92521, USA\\
       Department of Astronomy, National Research Institute of Astronomy and Geophysics, Helwan, 11421 Egypt}
\email{melha004@ucr.edu}

\author{Gillian Wilson}
\affil{Department of Physics and Astronomy, University of California Riverside, 900 University Avenue, Riverside, CA 92521, USA}

\author{Anatoly Klypin}
\affil{Astronomy Department, New Mexico State University, Las Cruces, NM 88001, USA}

\begin{abstract}
We introduce GalWeight, a new technique for assigning galaxy cluster membership. This technique is specifically designed to simultaneously maximize the number of {\it bona fide} cluster members while minimizing the number of contaminating interlopers. The GalWeight technique can be applied to both massive galaxy clusters and poor galaxy groups. Moreover, it is effective in identifying members in both the virial and infall regions with high efficiency. 
We apply the GalWeight technique to MDPL2 \& Bolshoi N-body simulations, and find that it is $> 98\%$ accurate in correctly assigning cluster membership. We show that GalWeight compares very favorably against four well-known existing cluster membership techniques (shifting gapper, den Hartog, caustic, SIM). We also apply the GalWeight technique to a sample of twelve Abell clusters (including the Coma cluster) using observations from the Sloan Digital Sky Survey. We end by discussing GalWeight's potential for other astrophysical applications.
\end{abstract}

\section{Introduction}


The problem of contamination of kinematic samples of galaxies in clusters by foreground and background galaxies is longstanding. It arises because of the fact that only the projected positions and velocities of galaxies are measured in redshift surveys. Due to the lack of knowledge about the motion perpendicular to the line of sight, it is difficult to judge a priori which of the galaxies found close to a cluster in projected space are actually bound to it and a good tracer of the underlying potential. Excluding fiducial members or including unbound galaxies, or interlopers, may lead to significantly
incorrect estimates of the cluster mass.

Several methods have been suggested in the literature to address this problem. All these methods aim at cleaning the galaxy sample by removing non-members before attempting a dynamical analysis of the cluster. Some algorithms utilize only the redshift information, such as (i) the 3$\sigma$-clipping method \citep{Yahil77} which iteratively eliminates interlopers with velocities greater than 3$\sigma$;  (ii) the fixed gapper technique \citep{Beers90,Zabludoff90} in which any galaxy that is separated by more than a fixed value (e.g., $1\sigma$ of the sample or 500-1000 km s$^{-1}$) from the central body of the velocity distribution is rejected as a non-member; or (iii) the jackknife technique \citep{Perea90} which removes the galaxy whose elimination causes the largest change in the virial mass estimator. These methods are primarily based on statistical rules and some selection criteria. Other algorithms utilize both position and redshift information, such as (i) the shifting gapper technique \citep{Fadda96} which applies the fixed gapper technique to a bin shifting along the distance from the cluster center, or (ii) the \citet{denHartog96} technique that estimates the maximum (escape) velocity as a function of distance from the cluster center calculated either by the virial or projected mass estimator (e.g., \citealp{Bahcall81,Heisler85}).

In addition to the techniques described above, the spherical infall models (hereafter referred to as SIMs, e.g., \citealp{Gunn72,Yahil85,Regos89,Praton94}) can determine the infall velocity as a function of distance from the cluster center. The SIM in phase-space has the shape of two trumpet horns glued face to face \citep{Kaiser87} which enclose the cluster members. However, studies shows that clusters are not well fit by SIMs in projected phase-space diagram, because of the random motion of galaxies in the cluster outer region caused by the presence of substructure or ongoing mergers \citep{vanHaarlem93,Diaferio99}. A recent investigation \citep{Abdullah13} showed that SIMs can be applied to a sliced phase-space by taking into account the distortion of phase-space due to transverse motions of galaxies with respect to the observer and/or rotational motion of galaxies in the infall region in the cluster-rest frame. However, that is out of the scope of the current paper.

Another sophisticated method is the caustic technique described by \cite{Diaferio99} which, based on numerical simulations \citep{Serra13}, is estimated to be able  to identify cluster membership with  $\sim 95\%$ completeness within $3r_v$ ($r_v$ is the virial radius defined below). The caustic technique depends on applying the two-dimensional adaptive kernel method (hereafter, 2DAKM, e.g., \citealp{Pisani93,Pisani96}) to galaxies in phase-space ($R_p$, $v_z$), with the optimal smoothing length $h_{opt} = (6.24/N) \sqrt{(\sigma_{R_p}^2+\sigma_{v_z}^2)/2}$, where $\sigma_{R_p}$ and $\sigma_{v_z}$ are the standard deviations of projected radius and line-of-sight velocity, respectively, and $N$ is the number of galaxies. $\sigma_{R_p}$ and $\sigma_{v_z}$ should have the same units and therefore the coordinates ($R_p$, $v_z$) should be rescaled such that $ q = \sigma_v / \sigma_{R_p}$, where $q$ is a constant which is usually chosen to be 25 or 35 (additional details about the application of this technique may be found in \citealp{Serra11}). 

One more technique that should be mentioned here is the halo-based group finder \citep{Yang05,Yang07}. 
\citet{Yang07} were  able to recover true members with $\sim 95\%$ completeness in the case of poor groups ($\sim10^{13} \mbox{M}_\odot$). However, they found that the completeness dropped to $\sim 65\%$ for rich massive clusters ($\sim10^{14.5} \mbox{M}_\odot$). Also, theirs is an iterative method which needs to be repeated many times to obtain reliable members. Moreover, its application depends on some assumptions and empirical relations to identify the group members.

This paper introduces a simple and effective new technique to constrain cluster membership which avoids some issues of other techniques e.g., selection criteria, statistical methods, assumption of empirical relations, or need for multiple iterations. The paper is organized as follows. The simulations used in the paper are described in \S \ref{sec:sims}. In \S \ref{sec:Tech} the GalWeight technique is introduced and its efficiency at identifying \emph{bona fide} members is tested on MultiDark N-body simulations. In \S \ref{sec:comp}, we compare GalWeight with four well-known existing cluster membership techniques (shifting gapper, den Hartog, caustic, SIM). We apply GalWeight to twelve Abell clusters (including the Coma cluster) in \S \ref{sec:sdss}, and present our conclusions in  \S \ref{sec:conc}.
Throughout this paper we adopt $\Lambda$CDM with $\Omega_m=0.3$, $\Omega_\Lambda=0.7$, and $H_0=100$ $h$ km s$^{-1}$ Mpc$^{-1}$, $h = 1$.
\section{Simulations}\label{sec:sims}

In this section we describe the simulated data that we use in this work in order to test the efficiency of the GalWeight technique to recover the true membership of a galaxy cluster.

{\bf 1. MDPL2}: The MDPL2 \footnote{https://www.cosmosim.org/cms/simulations/mdpl2/} simulation is an N-body simulation of $3840^3$ particles in a box of co-moving length 1 $h^{-1}$ Gpc, mass resolution of $1.51 \times 10^9$ $h^{-1}$ M$_{\odot}$, and gravitational softening length of 5 $h^{-1}$ kpc (physical) at low redshifts from the suite of MultiDark simulations (see Table 1 in \citealp{Klypin16}). It was run using the L-GADGET-2 code, a version of the publicly available cosmological code GADGET-2  \citep{Springel05}. It assumes a flat $\Lambda$CDM cosmology, with cosmological parameters $\Omega_\Lambda$ = 0.692, $\Omega_m$ = 0.307, $\Omega_b$ = 0.048, $n$ = 0.96, $\sigma_8$ = 0.823, and $h$ = 0.678 \citep{Planck14}. MDPL2  provides a good compromise between numerical resolution and volume \citep{Favole16}. It also provides us with a large number of clusters of different masses extended from $0.7\times10^{14}$ to $37.4\times10^{14}$ $h^{-1}$ $M_{\odot}$ (the range used to test the efficiency of GalWeight).

{\bf 2. Bolshoi}: The Bolshoi simulation is an N-body simulation of $2048^3$ particles in a box of co-moving length 250 $h^{-1}$ Mpc, mass resolution of $1.35 \times 10^8$ $h^{-1}$ M$_{\odot}$, and gravitational softening length of 1 $h^{-1}$ kpc (physical) at low redshifts. It was run using the Adaptive Refinement Tree (ART) code \citep{Kravtsov97}. It assumes a flat $\Lambda$CDM cosmology, with cosmological parameters ($\Omega_\Lambda$ = 0.73, $\Omega_m$ = 0.27, $\Omega_b$ = 0.047, $n$ = 0.95, $\sigma_8$ = 0.82, and $h$ = 0.70. Bolshoi provides us with clusters of higher mass resolution than MDPL2. Thus, we use both simulations to test the efficiency of GalWeight to recover the true membership.

For both simulations halos are identified using the Bound Density Maximum (BDM) algorithm \citep{Klypin97,Riebe13}, that was extensively tested (e.g., \citealp{Knebe11}) which identifies local density maxima, determines a spherical cut-off for the halo with overdensity equal to 200 times the critical density of the Universe ($\rho = 200 \rho_c$) for MDPL2 and 360 times the background matter density of the Universe ($\rho = 360 \rho_{bg}$), and removes unbound particles from the halo boundary. Among other parameters, BDM provides a virial masses and radii.   The virial mass is defined as $M_{v} =\frac{4}{3} \pi 200 \rho_c r_{v}^3$ for MDPL2 and  $M_{v} =\frac{4}{3} \pi 360 \rho_{bg} r_{v}^3$ for Bolshoi (see \citealp{Bryan98,Klypin16}). The halo catalogs are complete for halos with circular velocity $v_c \geq 150$ km s$^{-1}$ for MDPL2 \citep{Klypin16} and $v_c \geq 100$ km s$^{-1}$ for Bolshoi (e.g., \citealp{Klypin11,Busha11,Old15}).


For both MDPL2 and Bolshoi the phase-space (line-of-sight velocity $v_z$ versus projected radius $R_p$) of a distinct halo (cluster) is constructed as follows. 
We assume the line-of-sight to be along the z-direction and the projection to be on the x-y plane. We select a distinct halo of coordinates ($x^h,y^h,z^h$) and velocity components ($v_x^h,v_y^h,v_z^h$), and then we calculate the observed line-of-sight velocity of a subhalo, taking the Hubble expansion into account, as $v_{z} = (v_z^g- v_z^h) + H_0 (z^g-z^h)$, where ($x^g,y^g,z^g$) and ($v_x^g,v_y^g,v_z^g$) are the coordinates and velocity components of the subhalo, respectively. Finally, we select all subhalos within a projected radius of \footnote{Throughout the paper we utilize small $r$ to refer to 3D radius and capital $R$ to refer to projected radius.}$R_{p,max} = 10$ $h^{-1}$ Mpc from the center of distinct halo and within a line-of-sight velocity interval of $|v_{z,max}| = 3500$ km $\mbox{s}^{-1}$. These values are chosen to be sufficiently large to exceed both the turnaround radius and the length of the Finger-of-God (hereafter, FOG) which are typically $\sim7-8~h^{-1}$ Mpc and $\sim 6000$ km s$^{-1}$ respectively for massive clusters. The turnaround radius $r_t$ is the radius at which a galaxy's peculiar velocity ($v_{pec}$) is canceled out by the global Hubble expansion. In other words, it is the radius at which the infall velocity vanishes ($v_{inf} = v_{pec} - H~ r = 0$). 

\section{The Galaxy Weighting Function Technique (GalWeight)} \label{sec:Tech}

\begin{figure} \hspace*{-0.75cm}
\includegraphics[width=14cm]{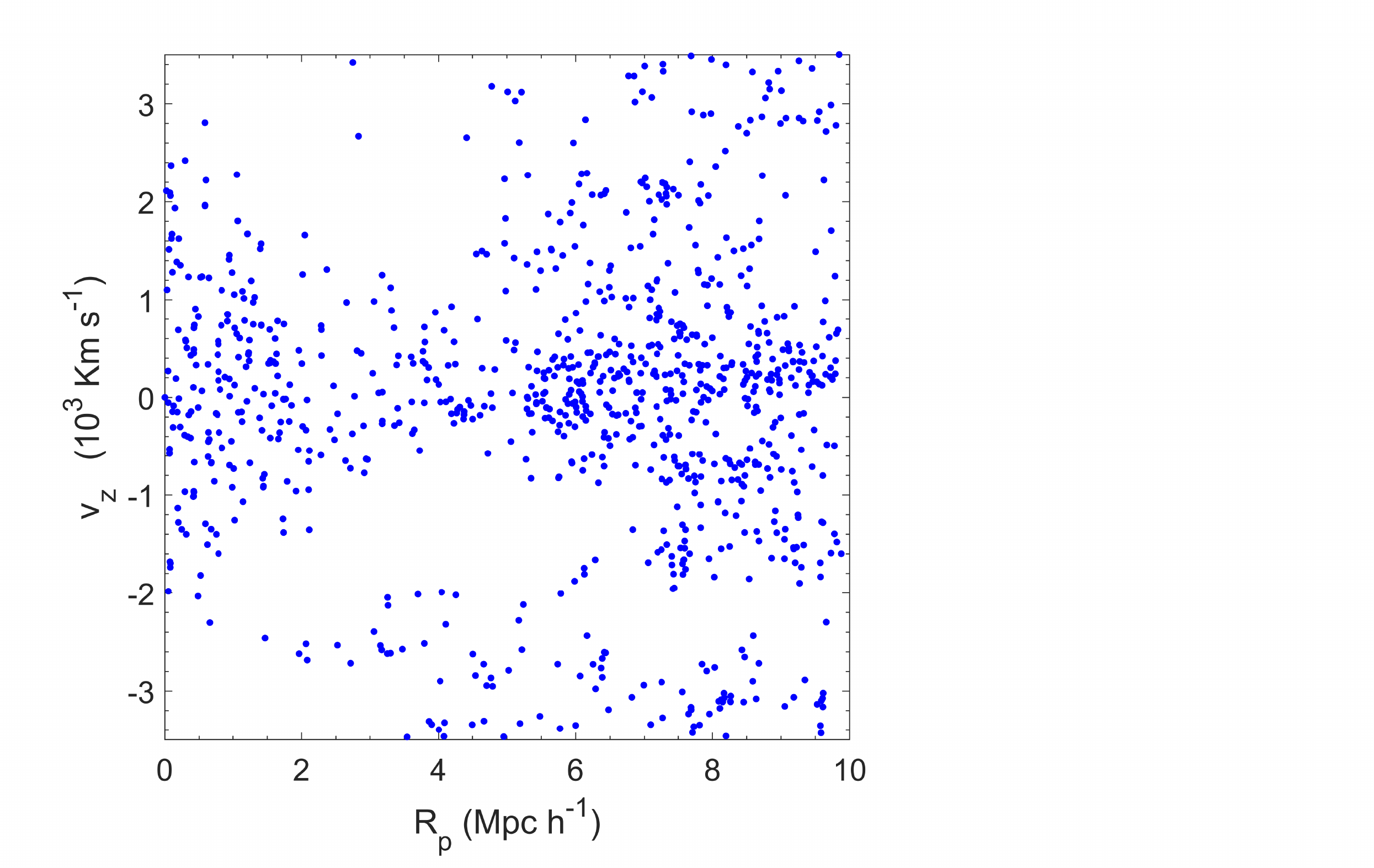} \vspace{-0.5cm}
\caption{Line-of-sight velocity $v_z$ as a function of projected radius $R_p$ in the extended region around a simulated cluster of mass $9.37 \times 10^{14}$ $h^{-1}$ M$_{\odot}$ selected from the Bolshoi simulation. The Finger-of-God is clearly seen in the main body of the cluster within $R_p\lesssim 1$ Mpc $h^{-1}$. The effect of the mass concentration in and around the cluster is manifested as a concentration of galaxies around $v_z=0$ line well outside the cluster itself. Interlopers are mostly galaxies at large projected distances and large peculiar velocities. In \S~\ref{sec:Tech} and in Figures~\ref{fig:F02C}, \ref{fig:F03C} \& \ref{fig:totweight} we show in detail how  GalWeight can be applied to this cluster to distinguish between interlopers and cluster members (Figure~\ref{fig:example}).
}
\label{fig:F01C}
\end{figure}

In this section, we describe the GalWeight technique in detail and demonstrate its use by applying it interactively to  a simulated cluster of mass $9.37 \times 10^{14}$ $h^{-1}$ M$_{\odot}$ selected from the Bolshoi simulation. Figure~\ref{fig:F01C} shows the phase-space distribution  of subhalos (galaxies) near the center of the simulated cluster.

\begin{figure*} \hspace*{1.5cm}
\includegraphics[width=13.5cm]{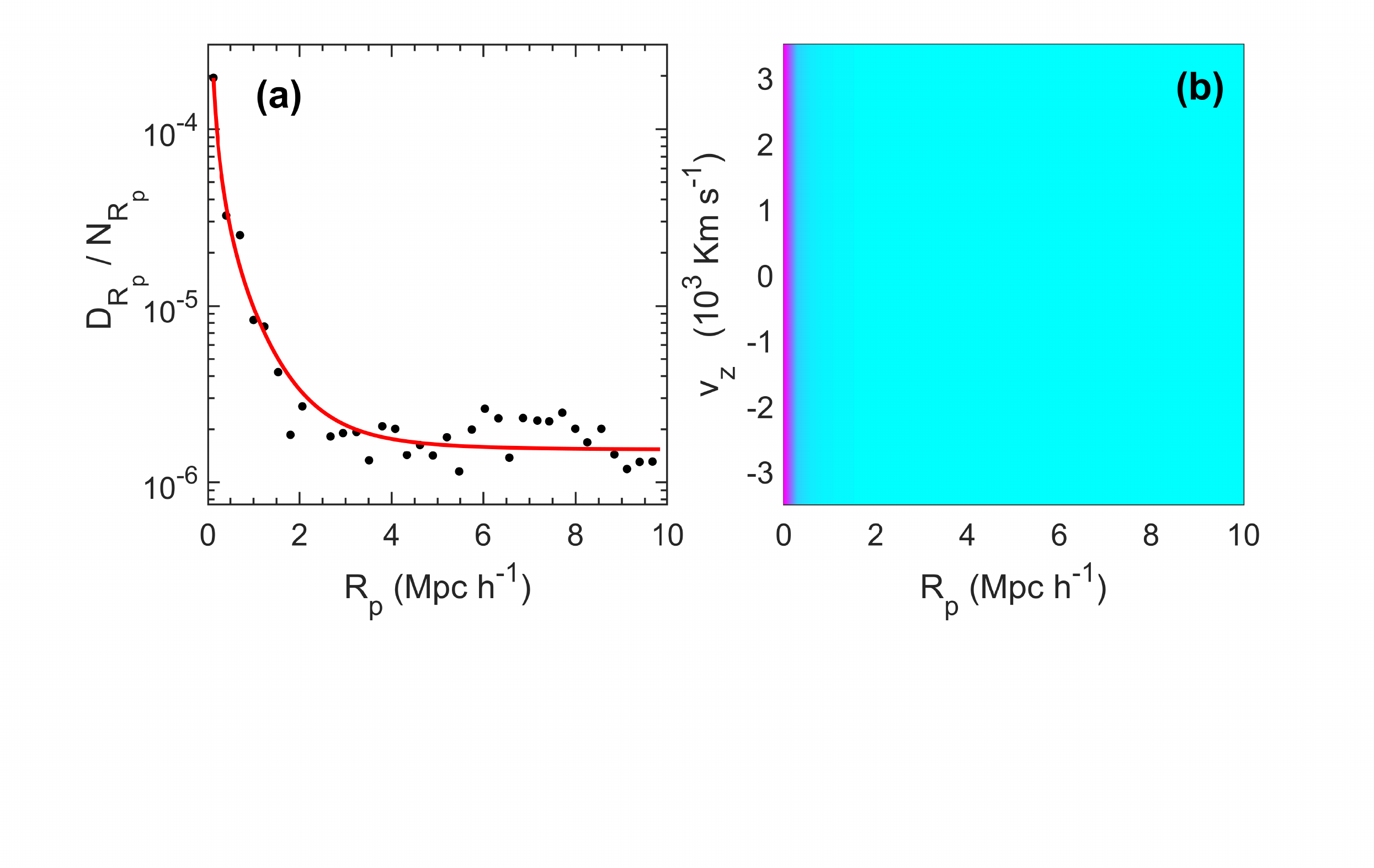} \vspace{-2.55cm}
\caption{Weighting function along projected radius $R_{p}$ for the simulated cluster of mass $9.37 \times 10^{14}$ $h^{-1}$ M$_{\odot}$ selected from Bolshoi (see \S~\ref{sec:Prob} A.1). The left panel (a) shows the function $\mathcal{D}_{R_p}$ derived from the data (black points, Equation (\ref{eq:ProbR})), normalized by Equation (\ref{eq:ProbRN}), and fitted by $\mathcal{W}_{R_p}$ (red curve, Equation (\ref{eq:king})). The right panel (b) presents its corresponding probability density function in phase-space diagram. As shown in (a \& b), the weighting is greatest at $R_p = 0$ and decreases outwards.
}
\label{fig:F02C}
\end{figure*}

\begin{figure*} \hspace*{1.5cm} 
\includegraphics[width=13.5cm]{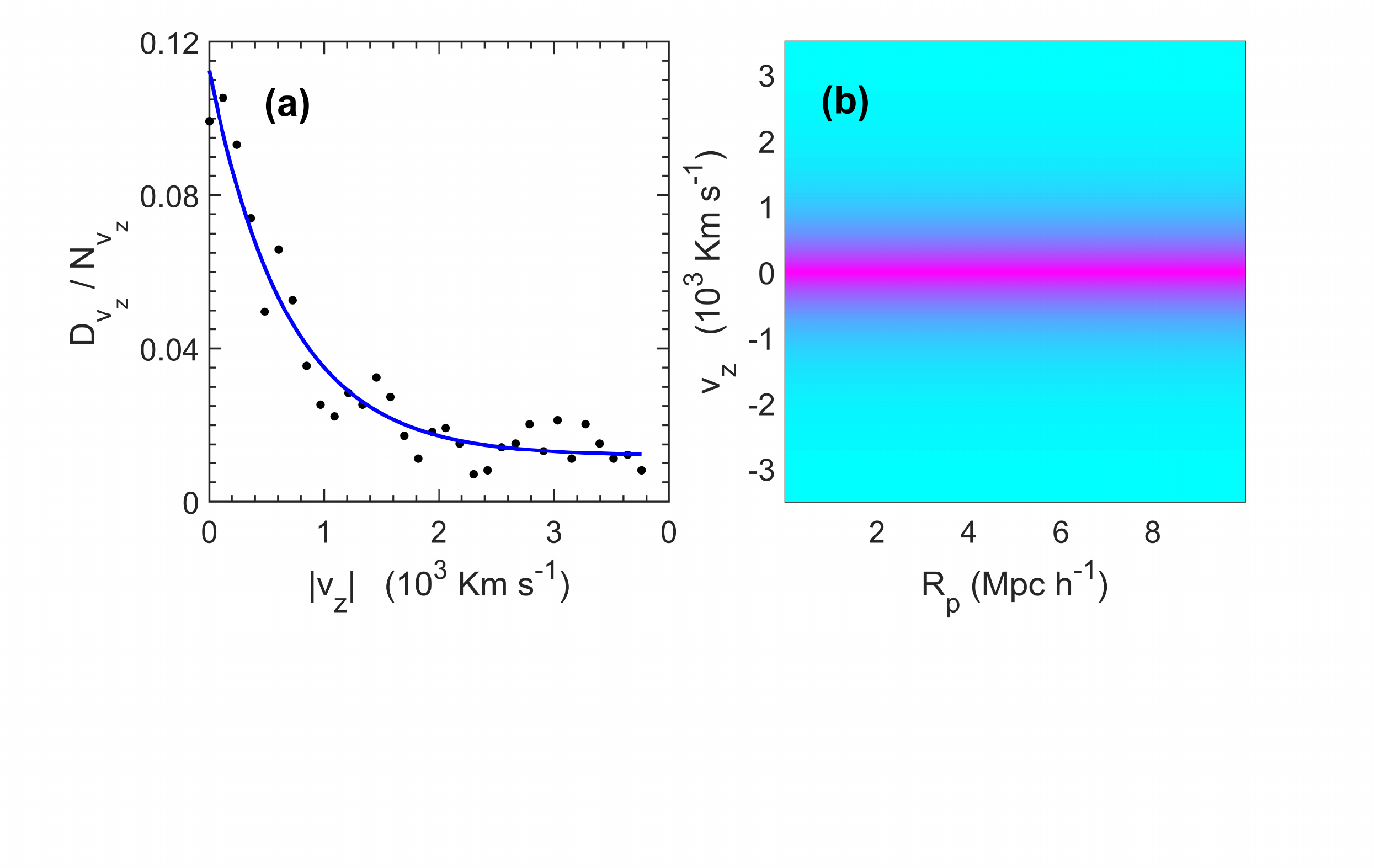} \vspace{-2.55cm}
\caption{Weighting function along line-of-sight velocity $v_{z}$ for the simulated cluster selected from Bolshoi. The left panel (a) shows the function $\mathcal{D}_{v_z}$ calculated from the data (black points, Equation  (\ref{eq:ProbV})), normalized by Equation (\ref{eq:ProbVN}), and fitted by $\mathcal{W}_{v_z}$ (blue curve, Equation (\ref{eq:Exp})). The right panel (b) presents its corresponding probability density function in phase-space. As shown in (a \& b), the applied weight is greatest at $v_z =0$ and decreases as the absolute line-of-sight velocity increases.}
\label{fig:F03C}
\end{figure*}

The GalWeight technique works by assigning a weight to each galaxy $i$ according to its position ($R_{p,i}$,$v_{z,i}$) in the phase-space diagram. This weight is the product of two separate two-dimensional weights which we refer to as the {\bf{dynamical}} and  {\bf{phase-space}} weights. The dynamical weight (see \S~\ref{sec:Prob} parts A.1 and A.2, and Figure~\ref{fig:totweight}a which is the product of Figure~\ref{fig:F02C}b and Figure~\ref{fig:F03C}b) is calculated from the surface number density $\Sigma(R_p)$,  velocity dispersion $\sigma_{v_z}(R_p)$, and standard deviation $\sigma_{R_p}(v_z)$ profiles of the cluster. The phase-space weight (see \S~\ref{sec:Prob} part B and Figure~\ref{fig:totweight}b) is calculated from the two-dimensional adaptive kernel method that estimates the probability density underlying the data and consequently identification of clumps and substructures in the phase-space (\citealp{Pisani96}).
The total weight is then calculated as the product of the dynamical and phase-space weights (see \S~\ref{sec:Prob} part C and Figure~\ref{fig:totweight}c). The advantage of using the total weight rather than the  dynamical weight or the phase-space weight alone is discussed in \S~\ref{sec:weights}. 

\subsection{Galaxy Weighting Functions}  \label{sec:Prob}

\noindent {\bf A. Dynamical Weighting $\mathcal{W}_{dy}(R_p,v_z)$} 
 
In calculating the dynamical weighting function, we assume that the weighting we apply should be larger at the cluster center i.e., at the origin in phase-space (Figure~\ref{fig:F01C}), and decay along both the $R_p$ and $v_z$ phase space axes. This weighting function is, therefore, a product of two individual weighting functions; one which decays along the $R_p$-axis and the other along the $v_z$-axis as described below.

\noindent {\bf A.1. $R_p$-axis Weighting Function, $\mathcal{W}_{R_p} (R_p)$}

In order to calculate the projected radius weighting function, $\mathcal{W}_{R_p} (R_p)$, we select two properties that are strongly correlated with  projected radius and with the dynamical state of a cluster. 

The first property is the {\bf\ Surface Number Density Profile $\Sigma(R_p)$}, defined as the number of galaxies per unit area as a function of distance from the cluster center. It has its maximum value at the cluster center and decreases with radial distance, and is also strongly correlated with the mass distribution of the cluster.  The significance of introducing $\Sigma(R_p)$ for calculating $\mathcal{W}_{R_p}$ is that the velocities of member galaxies in the core of some clusters can be as high as $\approx 3000$ km s$^{-1}$. It produces the Kaiser or FOG effect (see \citealp{Kaiser87}). This FOG distortion is the main reason that many membership techniques fail to correctly identify galaxies in the core with high line-of-sight velocities as members. Thus, $\Sigma(R_p)$ is essential to recover the members in the cluster core. In other words, ignoring $\Sigma(R_p)$ means missing some of the cluster members in the core.

The second property is the {\bf Projected Velocity Dispersion Profile, $\sigma_{v_z}(R_p)$}. The significance of introducing $\sigma_{v_z}(R_p)$ for calculating $\mathcal{W}_{R_p}$ is that it characterizes the dynamical state of a cluster from its core to its infall region. Specifically, there are random motion of member galaxies in the infall region due to the presence of substructures and recent mergers (e.g., \citealp{vanHaarlem92,vanHaarlem93,Diaferio97}). This effect of random motion can be taken into account implicitly in $\sigma_{v_z}(R_p)$. This is the main reason why the SIM technique fails in the cluster outskirts in the projected phase-space. Thus, $\sigma_{v_z}(R_p)$ is essential to recover the members in the cluster infall region. In other words, ignoring $\sigma_{v_z}(R_p)$ means missing some of the cluster members in the infall region.

Thus, the weighting $\mathcal{W}_{R_p}(R_p)$ in the projected radius direction can be calculated by introducing the function  $\mathcal{D}_{R_p}(R_p)$ that is given by

  \begin{equation} \label{eq:ProbR}
   \mathcal{D}_{R_p} (R_p) = \frac{\Sigma(R_p) \sigma_{v_z}(R_p)}{R_p^\nu},
  \end{equation}
\noindent with the normalization 

  \begin{equation} \label{eq:ProbRN}
   \mathcal{N}_{R_p} = \int_{0}^{R_{p,max}} \mathcal{D}_{R_p} (R_p) dR_p,
  \end{equation}

\noindent where $R_{p,max}$ is the maximum projected radius in phase-space. 
The denominator $R_p^\nu$, where the slope of the power low $\nu$ is a free parameter in the range $-1 \lesssim \nu \lesssim 1$, is introduced in Equation (\ref{eq:ProbR}) to provide flexibility and generalization for the technique.
The free parameter $\nu$ is selected to adjust the effect of the distortion of FOG in the core and the distortion of the random motion in the outer region. It is defined as 

\begin{equation}
\nu =\frac{\sigma_{FOG}(R\leq0.25)}{\sigma_{rand}(0.25<R\leq4))} -1, 
\end{equation}

\noindent where $\sigma_{FOG}$ is the velocity dispersion of the core galaxies and $\sigma_{rand}$ is the velocity dispersion of the galaxies outside the core. 

The function $\mathcal{D}_{R_p} (R_p)$, calculated from the data, is contaminated by interlopers that cause scattering, especially at large projected distances (see black points in the left panel of Figure~\ref{fig:F02C}). Therefore, in order to apply a smooth weighting function, we fit $\mathcal{D}_{R_p} (R_p)$ with an analytical function. Any analytical function that is a good fit to $\mathcal{D}_{R_p} (R_p)$ could be utilized. In this paper we choose to use the function

  \begin{equation} \label{eq:king}
   \mathcal{W}_{R_p}(R_p)= \mathcal{A}_0\left(1+\frac{R_p^2}{a^2}\right)^{\gamma}+\mathcal{A}_{bg},
   \end{equation}

\noindent  which has four parameters: $a$ is a scale radius ($0 < a \lesssim 1$), $\gamma$ is a slope of the power law ($-2 \lesssim \gamma < 0$), and $\mathcal{A}_{0}$ and $\mathcal{A}_{bg}$ are the central and background weights in the $R_p$-direction. These parameters are determined by applying the chi-squared algorithm using the Curve Fitting MatLab Toolbox. Note that the analytical function we selected here has the same form as the generalized King model \citep{King72,Adami98}. 

Thus, the weight $\mathcal{W}_{R_p}(R_{p,i})$ of each galaxy can be calculated according to its projected radius $R_{p}$ from the cluster center. The weighting along $R_p$ is shown in Figure~\ref{fig:F02C}a, where the function $\mathcal{D}_{R_p} (R_p)$ is normalized using Equation (\ref{eq:ProbRN}). The data are  smoothed and approximated using Equation (\ref{eq:king}) (shown as red line). The right panel (b) shows the projected radius weight function in phase space.\\

\noindent{\bf A.2. $v_z$-axis Weighting Function, $\mathcal{W}_{v_z} (v_z)$}

In phase-space, most members are concentrated near the line $v_z = 0$ and the number of members decreases with increasing absolute line-of-sight velocity. The weighting function along $v_z$-axis can, therefore, be approximated by the histogram of the number of galaxies per bin, $N_{bin} (v_z)$, or equivalently  the standard deviation of projected radius, $\sigma_{R_p} (v_z)$, directed along the line-of-sight velocity axis, normalized by the total number of galaxies $N_{tot}$ in the cluster field. In other words, the weighting in the line-of-sight velocity direction can be calculated by introducing the function $\mathcal{D}_{v_z} (v_z)$ that is given by

   \begin{equation} \label{eq:ProbV}
   \mathcal{D}_{v_z} (v_z) = \sigma_{R_p} (v_z),
   \end{equation}

\noindent with the normalization 

  \begin{equation} \label{eq:ProbVN}
   \mathcal{N}_{v_z} = \int_{-v_{z,max}}^{v_{z,max}} \mathcal{D}_{v_z} (v_z) dv_z,
  \end{equation}

\noindent where $v_{z,max}$ is the maximum line-of-sight velocity of phase-space. As above, to obtain a smooth weighting function in $v_{z}$, the histogram or equivalently $\mathcal{D}_{v_z} (v_z)$ can be fitted by an analytical function. In this paper we select an exponential model that is given by\\

\begin{figure*}  \hspace*{0.5cm}
\includegraphics[width=21cm]{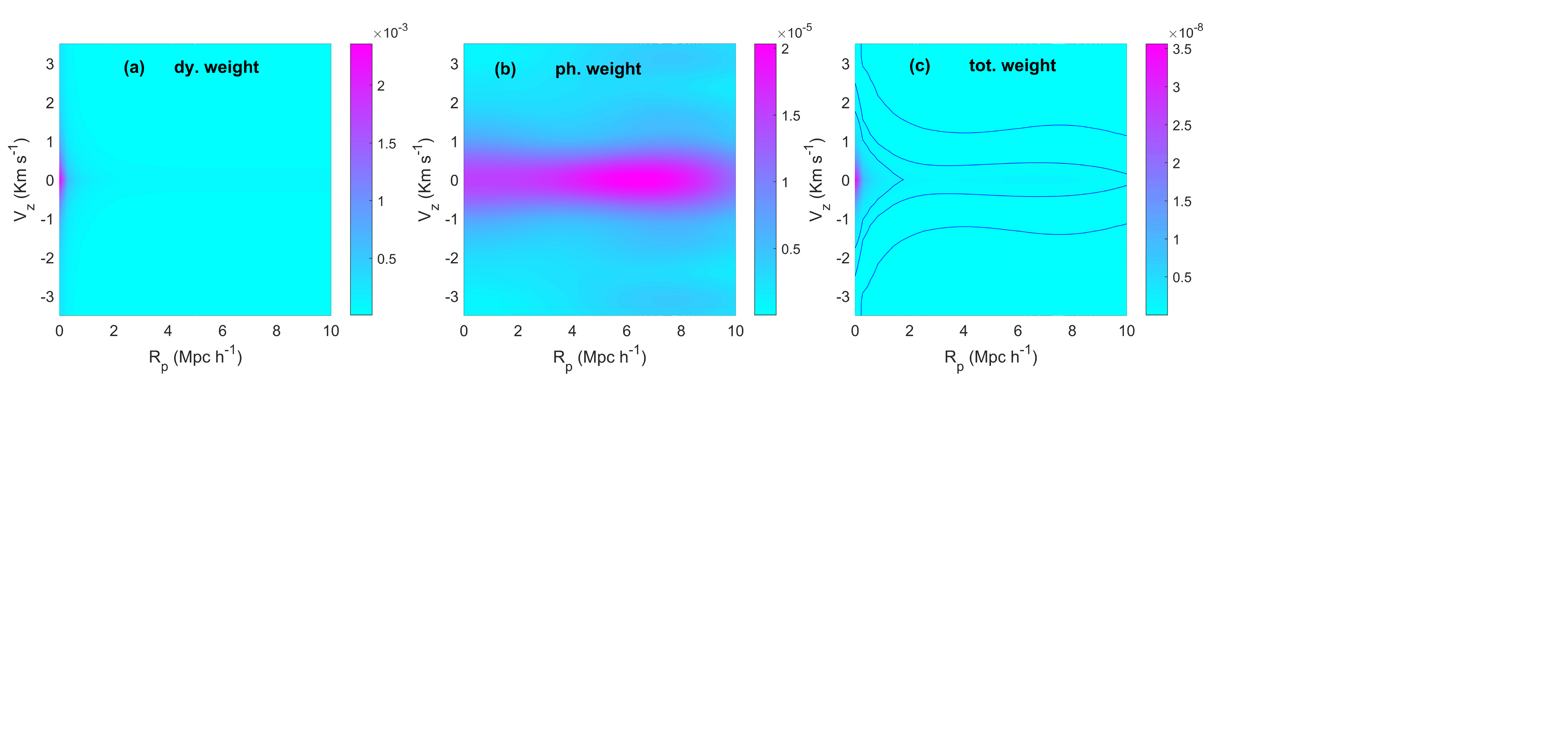} \vspace{-5.5cm}
\caption{Weights to be applied as a function of position in line-of-sight velocity/projected radius phase-space for the simulated cluster selected from the Bolshoi simulation. Panel (a) shows the dynamical weight $\mathcal{W}_{dy}$ (The product of the weights shown in Figures \ref{fig:F02C}b and \ref{fig:F03C}b). Panel (b) presents the phase-space weight $\mathcal{W}_{ph}$ calculated from the 2DAKM. The total weight $\mathcal{W}_{tot}  = \mathcal{W}_{dy} \times \mathcal{W}_{ph}$ is shown in panel (c) with explicitly drawing three contour weights. The weight $\mathcal{W}_{dy}$ is maximum at the origin (0,0) and decreases along both the $R_p$ and $v_z$ axes and $\mathcal{W}_{ph}$ gives higher weight for galaxy clumping around the center and substructures as well. Note that the scaling for each panel is independent, with magenta representing maximum values.}
\label{fig:totweight}
\end{figure*}

  \begin{equation} \label{eq:Exp}
  \mathcal{W}_{v_z} (v_z) = \mathcal{B}_0 \exp{(b \ v_z)}+\mathcal{B}_{bg},
  \end{equation}

\noindent where $\mathcal{B}_0$ is the central weight, $\mathcal{B}_{bg}$ is the background weight in $v_{z}$ and $b$ is scale parameter ($-0.01 \lesssim b < 0$). Again, these parameters are determined by applying the chi-squared algorithm using the Curve Fitting MatLab Toolbox. The weighting along $v_z$, is shown in Figure~\ref{fig:F03C}a, where the function $\mathcal{D}_{v_z} (v_z)$ (black points) is normalized using Equation (\ref{eq:ProbVN}). The data are smoothed and approximated by Equation (\ref{eq:Exp}) for an exponential model (blue curve). The right panel (b) shows the resulting exponential-model weight as a function of location in line-of-sight velocity/projected radius phase-space. As shown in (a \& b), the applied weight is greatest at $v_z =0$ and decreases as the absolute line-of-sight velocity increases.\\

We can now construct a two-dimensional {\bf{dynamical weight $\mathcal{W}_{dy}(R_p,v_z)$}} by multiplying $\mathcal{W}_{R_p} (R_p)$ and $\mathcal{W}_{v_z} (v_z)$ together:

\begin{equation} \label{eq:ProbDy}
   \mathcal{W}_{dy}(R_p,v_z)  = \mathcal{W}_{R_p} (R_p) \mathcal{W}_{v_z} (v_z),
  \end{equation}

\noindent $\mathcal{W}_{dy}(R_p,v_z)$ is shown in the left panel of Figure~\ref{fig:totweight}, and is the product of the weights shown in Figure~\ref{fig:F02C}b and Figure~\ref{fig:F03C}b. The weight is maximum at the origin, and decreases along both $R_p$ and $v_z$. 

To sum up, the dynamical weight is calculated from three properties (surface number density $\Sigma(R_p)$ and velocity dispersion $\sigma_{v_z}(R_p)$ along $R_p$, and standard deviation of projected radius $\sigma_{R_p} (v_z)$ along $v_z$) which are correlated strongly with the dynamics of the cluster. This weight takes into account the effects of the FOG in the cluster core and the random motion of galaxies in the infall region.

\noindent {\bf B. Phase-Space Weighting, $\mathcal{W}_{ph}(R_p,v_z)$}

This weighting is the coarse-grained phase-space density which can be simply calculated by the 2-dimensional adaptive kernel method (2DAKM, e.g., \citealp{Silverman86,Pisani96}). The kernel density estimator is the estimated probability density function of a random variable. For $N$ galaxies with coordinates $(x , y) = (R_p , v_z)$ the density estimator is given by

\begin{equation} \label{eq:kernal}
f(x,y)=\frac{1}{N} \sum_{i=1}^{N} \frac{1}{h_{x,i} h_{y,i}} K\left(\frac{x-X_i}{h_{x,i}}\right) K\left(\frac{y-Y_i}{h_{y,i}}\right)
\end{equation}

\noindent where, the kernel $K(t)$ is given by Gaussian distribution

\begin{equation} \label{eq:Gkernal}
K(t) = \frac{1}{\sqrt{(2\pi)}} \exp {\left(-\frac{1}{2} t^2\right)}
\end{equation}

\noindent and $h_{j,i} = \lambda_i h_j$ is the local bandwidth, $h_j = \sigma_{j} N^{-1/6}$ is the fixed bandwidth for 2-dimensional space and $\sigma_j$ is the standard deviation for j=\{x,y\}. The term $\lambda_i = \left[\gamma/f_0(x_i,y_i)\right]^{0.5}$ and $\log{\gamma}=\sum_i\log{f_0(x_i,y_i)}/N$, where $f_0(x_i,y_i)$ is given by Equation \ref{eq:kernal} for $\lambda_i=1$ (see also, \citealp{Shimazaki10}).
\begin{figure*} \hspace*{0.5cm}
\includegraphics[width=19cm]{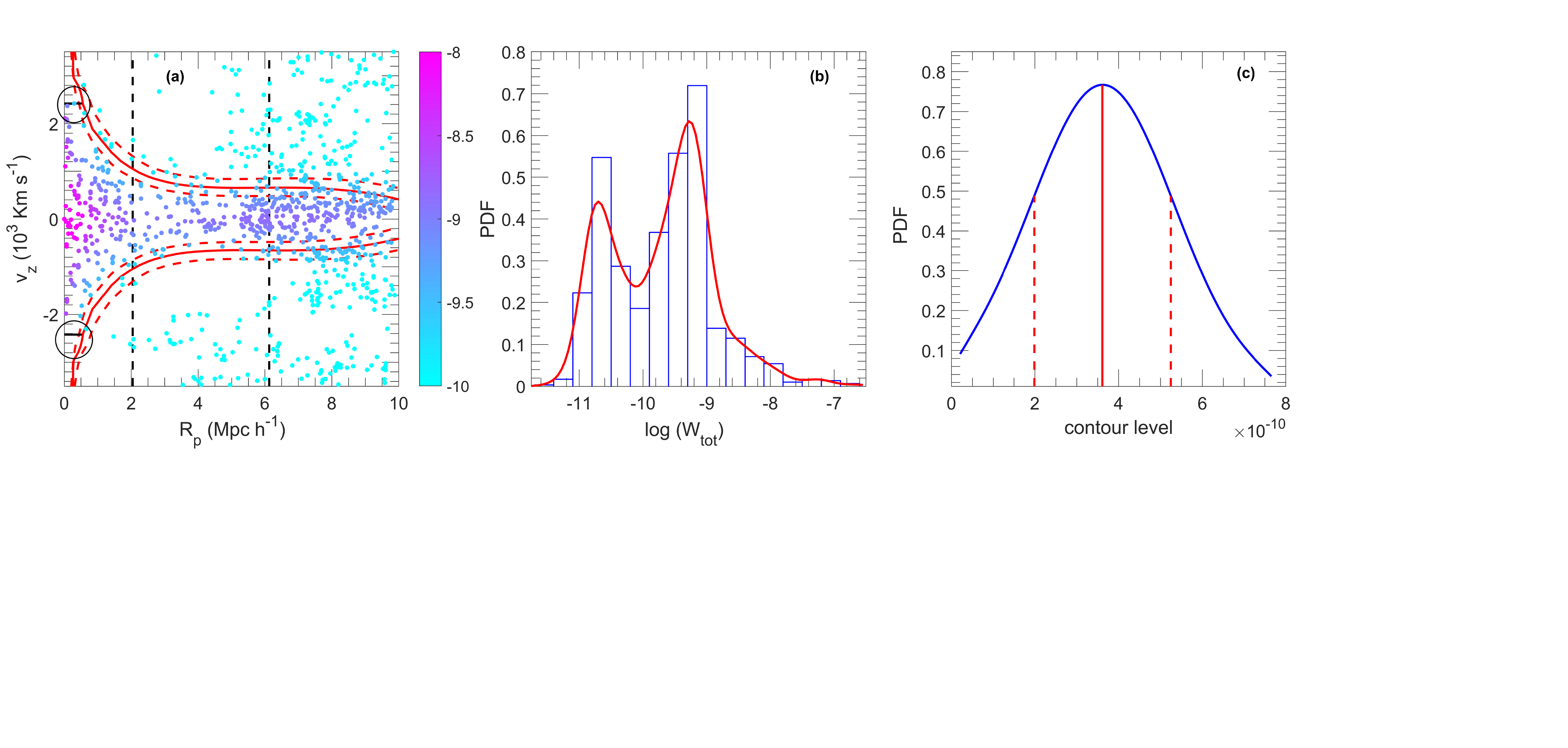} \vspace{-4cm}
\caption{Identification of the simulated cluster membership from weighted galaxies. Panel (a) shows the weight of each galaxy in line-of-sight velocity/projected radius phase-space (magenta color indicates higher weight). Panel (b) shows a histogram or PDF of the weight applied to each galaxy, $\mathcal{W}_{tot} (R_{p,i},v_{z,i})$. 1DAKM fitting returns a bimodal PDF in this example of the simulated cluster. We choose to use the number density method (NDM, \citealp{Abdullah13})  to identify the contour weight value which separates cluster members from interlopers. This is shown by the solid red vertical line in panel (c) and solid red line in panel (a). $1\sigma$ confidence intervals are shown by the two red dashed lines. The two vertical dashed-black lines represent the virial and turnaround radii, where the cluster members are those enclosed by the best contour line and within the turnaround radius. We impose one additional cut, shown by the black solid lines in panel (a), cutting the red contour line in the very inner radius by the maximum $v_z$ of the enclosed members.}
\label{fig:example}
\end{figure*}

Consequently, applying 2DAKM for the distribution of galaxies in the phase-space demonstrates high weights for positions of high-density distribution of galaxies. Therefore, the main purpose of introducing the phase-space weight is to take into account the effect of the presence of any clump or substructure in the field that cannot be counted by the dynamical weight. Also, the phase-space weight is introduced to reduce the excessive increase of dynamical weight near the center (see \S \ref{sec:weights}). The phase-space weight $\mathcal{W}_{ph}(R_p,v_z)$ is shown in Figure~\ref{fig:totweight}b 
that gives more weights for galaxies in clumps and substructures, and from the distribution of galaxies in the cluster field this weighting function is maximum around the cluster center.

\noindent {\bf C. Total Weighting, $\mathcal{W}_{tot}(R_p,v_z)$}

The total weighting function is calculated as

  \begin{equation} \label{eq:ProbTot}
   \mathcal{W}_{tot}(R_p,v_z)  = \mathcal{W}_{dy} (R_p,v_z) \mathcal{W}_{ph} (R_p,v_z),
  \end{equation}

\noindent and shown in Figure~\ref{fig:totweight}c for the simulated cluster. It shows the probability distribution function of the total weight $\mathcal{W}_{tot}(R_p,v_z)$. The weighting in Figure~\ref{fig:totweight}c is then applied to individual galaxies. 
Figure~\ref{fig:example}a shows Fig~\ref{fig:F01C} once again, but now after applying the ``total weighting". 
We still need to separate cluster members from interlopers.
We explain how to do that in \S~\ref{sec:mem}.

\subsection{Membership Determination} \label{sec:mem}

Figure~\ref{fig:example}a shows the weight of each galaxy in the simulated cluster phase-space. The question is now how to utilize the weighted galaxies in phase-space to best identify cluster members.
One would like to identify a single, optimal weight value which separates cluster members from field galaxies i.e., identify the best contour weight to select in panel (a). One way is to consider the probability distribution function (PDF), or histogram of the total weight for all galaxies, which is shown in Figure~\ref{fig:example}b. Fitting the PDF  using a 1DAKM reveals two obvious peaks (bimodal PDF). 
One might imagine simply drawing a vertical line to separate the members located on the right with higher weights from the interlopers located on the left. However, not all clusters show this bimodality in the PDF of $\mathcal{W}_{tot}$. Another way could be to exclude all galaxies that have weights less than, for example, $3\sigma$ from the average value of the main peak (i.e., $\mathcal{W}_{cut} = \mathcal{W}_{peak} - 3\sigma$). However, attempting to do the separation by either of these two ways is subjective.

Therefore, we prefer to select the optimal contour weight by utilizing the Number Density Method (hereafter, NDM), a technique which was introduced in \citet{Abdullah13}. The goal in applying the method here, is to find the optimal contour weight (or line) that returns the maximum number density of galaxies. In other words, we select a certain contour line (weight) and calculate its enclosed area and number of galaxies, $N_{\rm in}$ (which contribute positively), then account for the number of galaxies, $N_{\rm out}$ (which contribute negatively) located outside this contour line. Then, the number density of this contour line can be calculated by ($N_{\rm in} - N_{\rm out})/{\rm Area}$ (see figure 9 in \citealp{Abdullah13}). 

\begin{figure*} \hspace*{1cm}
\includegraphics[width=22cm]{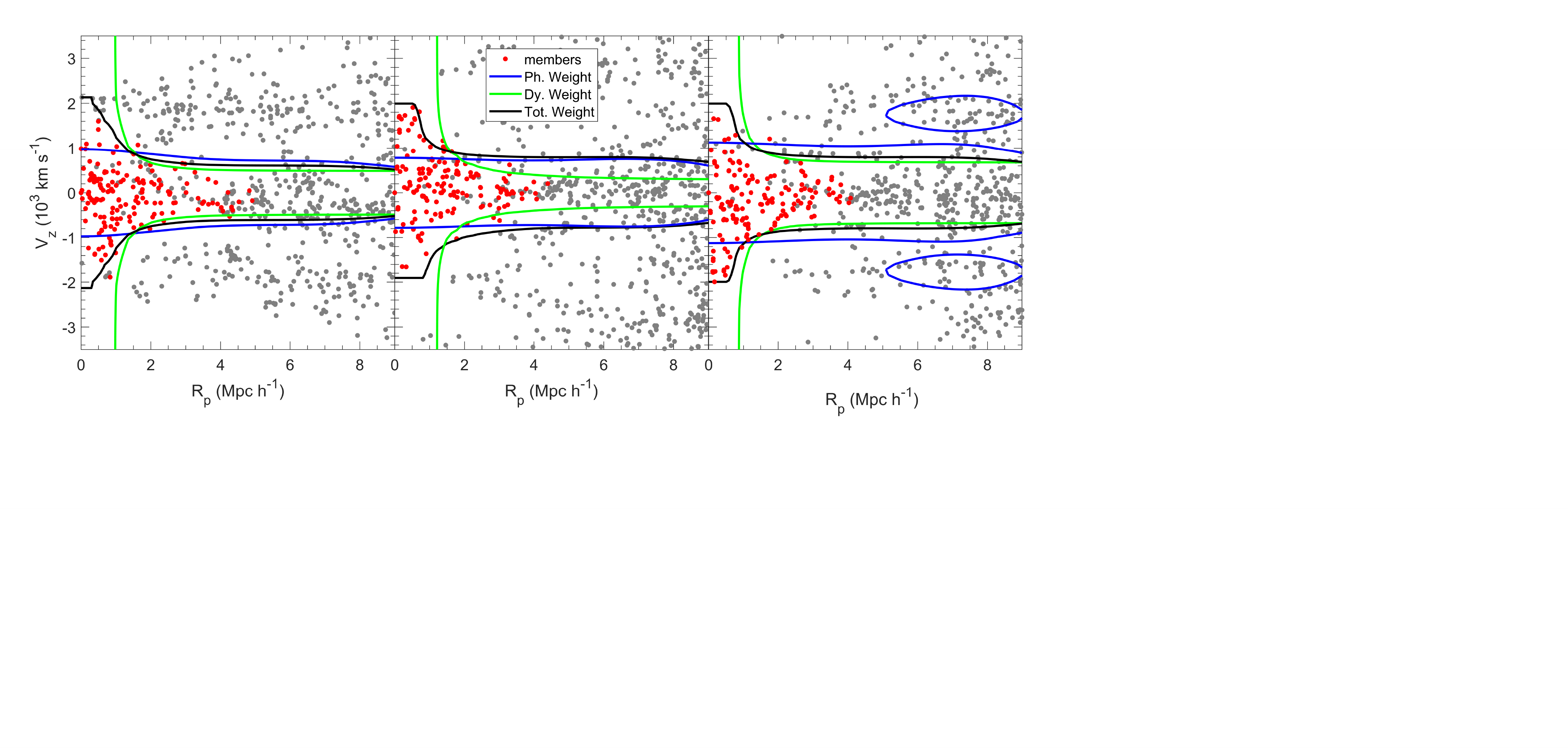} \vspace{-5.25cm}
\caption{Application of dynamical, phase-space, and total weights (green, blue, and black lines, respectively) to three simulated clusters taken from the Bolshoi simulation (\S~\ref{sec:sims}). The red points show true members within $3r_v$. Applying the dynamical weight alone (green) results in the inclusion of many galaxies within $R \sim$ 1 Mpc $h^{-1}$ with very high line-of-sight velocities. Applying the phase-space weight alone (blue), fails to recover some members in the core while simultaneously 
incorrectly including some interlopers at large distances due to the presence of nearby clusterings and clumps. The total weight (black), the product of the dynamical and phase-space weights, recovers true members effectively in both the core and infall regions (see Table \ref{tab:frac}).}
\label{fig:WeightsComp}
\end{figure*}

In Figure~\ref{fig:example}c the PDF of the number density of galaxies calculated by NDM is plotted for weights (contour lines) in the range $-12 \leq \log{\mathcal{W}_{tot}} \leq -6$. The optimal contour line corresponds to the maximum number density of galaxies, the value of weight which should be utilized as the separator of cluster members from interlopers, is shown by the red vertical solid line with $1\sigma$ confidence intervals shown by the red two vertical dashed lines. This optimal contour line with $1\sigma$ confidence  are shown as solid and dashed red lines in panel (a), respectively. 

As shown in Figure~\ref{fig:example}a the optimal contour line extends to large distances ($R \sim 10~ h^{-1}$ Mpc) and not all galaxies within this boundary are members. Therefore, the last step of GalWeight is to determine a cutoff radius within which the galaxies are assumed to be bounded. Thus, the cluster members are defined as the galaxies enclosed by the optimal contour line and within the cutoff radius. This cutoff radius can be adopted as the virial radius $r_v$ (which is the boundary of the virialized region) or the turnaround radius $r_t$ (which is the boundary of the cluster infall region). Note that the main goal of this paper is to introduce and test the efficiency of GalWeight to recover the true members in the virial and infall regions using simulations. Thus, knowing the virial radius of each simulated cluster we test the efficiency of GalWeight at $r_v$, $2r_v$, and $3r_v$ projected on the phase-space diagram as described in \S \ref{sec:simulations} and Table \ref{tab:frac} (see, e.g., \citealp{Serra13}). However, for our sample of the twelve Abell clusters (observations) $r_v$ and $r_t$ are determined from the mass profile estimated by the virial mass estimator and NFW mass profile \citep{NFW96,NFW97} as discussed in \S \ref{sec:sdss}.


We impose one additional cut, shown by the solid black lines highlighted by black circles in panel (a), to cut the red contour line in the very inner radius by the maximum $v_z$ of the enclosed members. This is because in some cases the optimal contour line extends to very high velocities in the innermost region ($R \lesssim 0.25 h^{-1}$ Mpc) without including any other members, so it is not necessarily to show this tail of the contour line.

The main steps in applying the GalWeight technique to determine cluster membership are summarized below:\\
{\bf{1.}} Make an appropriate cut in $R_p$ and $v_z$, and plot galaxies in line-of-sight velocity/projected radius phase-space.
In this paper, we use $R_{p,max} = 10$ $h^{-1}$ Mpc and $|v_{z,max}| = 3500$ km $\mbox{s}^{-1}$.\\
{\bf{2.}} Calculate the function $\frac{\Sigma(R_p) \sigma(R_p)}{R_p^\nu}$ and fit it with an analytical model (e.g., Equation~\ref{eq:king}) to obtain $\mathcal{W}_{R_p} (R_p)$.\\
{\bf{3.}} Calculate the function $\sigma_{v_z} (v_z)$ and fit it with an analytical model (e.g., Equation~\ref{eq:Exp}) to obtain $\mathcal{W}_{v} (v)$.\\
{\bf{4.}} Determine the dynamical weighting, $\mathcal{W}_{dy} (R_p,v_z) = \mathcal{W}_{R_p} (R_p) \times \mathcal{W}_{v_z} (v_z)$.\\
{\bf{5.}} Apply the 2DAKM in phase-space to determine the phase-space weighting, $\mathcal{W}_{ph} (R_p,v_z)$.\\
{\bf{6.}} Calculate the total weight $\mathcal{W}_{tot} (R_p,v_z) = \mathcal{W}_{dy} (R_p,v_z) \times \mathcal{W}_{ph} (R_p,v_z)$.\\
{\bf{7.}} Plot the PDF for all galaxy weights and apply a cut, retaining all galaxies with weight larger than this cut as members (NDM is used here to determine the optimal value of cut).\\
{\bf{8.}} Determine the cutoff radius ($r_v$ or $r_t$) using a dynamical mass estimator and identify cluster members as those enclosed by the optimal contour line and within the cutoff radius.
 
\subsection{Why do we use total weight rather than dynamical or phase-space weights?} \label{sec:weights}

One may ask why we depend on the total weight to assign a cluster membership rather than using the dynamical weight or phase-space weight alone. We present Figure~\ref{fig:WeightsComp} to help answer this question. It shows the phase-space of three Bolshoi simulated clusters (see \S \ref{sec:sims}). Using simulated clusters  brings the advantage that true members are known definitively. Figure~\ref{fig:WeightsComp} shows the optimal contour lines  determined by applying, separately, the dynamical weight (green line), the phase-space weight (blue line) and the total weight (black line). The red points show true members within $3r_v$.

\begin{table*} \centering
\caption{Efficiency of the GalWeight technique\\
determined by calculating $f_c$ and $f_i$ at $r_v$, $2r_v$ and $3r_v$ for a sample of $\sim$ 3000 clusters from the MDPL2 \& Bolshoi simulations.}
\label{tab:frac}
\begin{tabular}{cccccccccc}\hline
Mass Range&mean&number of&\multicolumn{3}{c}{$f_c$ }&&\multicolumn{3}{c}{$f_i$}\\
\multicolumn{2}{c}{($10^{14}~ h^{-1}~ \mbox{M}_\odot$)}&halos &$r_v$ & $2r_v$ & $3r_v$& & $r_v$ & $2r_v$ & $3r_v$\\
(1)&(2)&(3)&(4)&(5)&(6)&&(7)&(8)&(9)\\
\hline
\multicolumn{10}{c} {MDPL2}\\
\hline
0.73-37.39&4.28&1500 (All)&$0.993\pm0.014$ &$0.986\pm0.015$ &$0.981\pm0.013$&&$0.112\pm0.035$&$0.096\pm0.048$&$0.113\pm0.051$\\
\hline
0.73-2.00&1.37&253&  $0.998\pm0.050$ &$0.992\pm0.016$ &$0.981\pm0.018$&&$0.096\pm0.039$&$0.098\pm0.050$&$0.118\pm0.053$\\
2.00-4.00&3.16&617&  $0.993\pm0.015$ &$0.983\pm0.016$ &$0.979\pm0.012$&&$0.113\pm0.034$&$0.099\pm0.050$&$0.118\pm0.053$\\
4.00-8.00&5.37&484&  $0.989\pm0.016$ &$0.984\pm0.013$ &$0.982\pm0.011$&&$0.118\pm0.032$&$0.099\pm0.045$&$0.117\pm0.049$\\
8.00-37.39&11.20&146&$0.988\pm0.013$ &$0.988\pm0.010$ &$0.988\pm0.013$&&$0.121\pm0.028$&$0.105\pm0.043$&$0.122\pm0.045$\\
\hline
\hline
\multicolumn{10}{c} {Bolshoi}\\
\hline
0.70-10.92&1.53&500 (1500) (All)&$0.995\pm0.011$ &$0.981\pm0.021$ &$0.971\pm0.020$&&$0.126\pm0.045$&$0.217\pm0.109$&$0.226\pm0.102$\\
\hline
0.70-2.00&1.31&415 (1194)&  $0.996\pm0.011$ &$0.983\pm0.0208$ &$0.972\pm0.019$&&$0.124\pm0.047$&$0.218\pm0.109$&$0.227\pm0.102$\\
2.00-4.00&2.70&72 (252)&  $0.992\pm0.012$ &$0.975\pm0.023$ &$0.967\pm0.025$&&$0.133\pm0.040$&$0.128\pm0.103$&$0.227\pm0.105$\\
4.00-8.00&4.43&11 (48)&  $0.990\pm0.012$ &$0.970\pm0.022$ &$0.961\pm0.022$&&$0.131\pm0.039$&$0.207\pm0.113$&$0.217\pm0.103$\\
8.00-10.92&9.68&2 (6)&$0.997\pm0.004$ &$0.982\pm0.024$ &$0.973\pm0.025$&&$0.130\pm0.018$&$0.270\pm0.116$&$0.250\pm0.094$\\
\hline
\end{tabular}
\begin{tablenotes}
\item $f_c$ is the completeness  or the fraction of the number of fiducial members identified by GalWeight as members relative to the actual number of members. 
\item $f_i$ is  the contamination or the fraction of interlopers incorrectly assigned to be members.  
\item 
Columns: (1) Cluster mass range; (2) cluster mean mass per bin; (3) the actual number of clusters per bin in simulations and the number between brackets gives the number of clusters in different orientations for Bolshoi; (4-6) and (7-9) are $f_c$ and $f_i$ for each mass bin at $r_v$, $2r_v$, and $3r_v$, respectively.
\end{tablenotes}
\end{table*}

In Figure~\ref{fig:WeightsComp}, the dynamical weight $\mathcal{W}_{dy}(R_p,v_z)$ (green; see also Figure~\ref{fig:totweight}a) is seen to be very smooth and idealised. In other words, $\mathcal{W}_{dy}(R_p,v_z)$ describes well an isolated galaxy cluster in phase-space. It does not take into account the effects of nearby clusters, clumps and/or substructures. Also, it shows an excessive increase near the cluster center ($\sim 1~ h^{-1}$~Mpc) and incorrectly includes interlopers near the center which have very high velocities. This effect is due to introducing $\Sigma(R_p)$ in $\mathcal{W}_{dy}(R_p,v_z)$, where the surface number density is very high near the cluster center. However, ignoring $\mathcal{W}_{dy}(R_p,v_z)$ leads to missing some cluster members especially those that close to the center in phase-space. Thus, $\mathcal{W}_{dy}(R_p,v_z)$ cannot be used on its own to assign cluster membership, but it is very important for correctly identifying members  with high line-of-sight velocities.

Figure~\ref{fig:WeightsComp} demonstrates that, on its own, phase-space weighting $\mathcal{W}_{ph}(R_p,v_z)$ also has some difficulty in recovering true cluster members (blue; see also Figure~\ref{fig:totweight}b). This is because it does not take into account the FOG effect in the cluster core, where those members that have high velocities do not have high concentration, so they are assigned low weights in phase-space and not counted as members. Also,  the presence of nearby clusterings and substructures have the effect of widening the ``optimal'' contour line. Consequently, it is very difficult to separate true members from galaxies (interlopers) located in nearby clumps. This results in the inclusion of some interlopers in the infall region. In summary, using $\mathcal{W}_{ph}(R_p,v_z)$ alone, simultaneously excludes some true members near the cluster center and includes some interlopers in the infall region.

We have shown that both the dynamical weight and phase-space weight have issues in identifying true members when applied alone. However, as the black solid line in Figure~\ref{fig:WeightsComp} shows, the total weight (the product of the dynamical and phase-space weights), is very effective. It can simultaneously identify  cluster members moving with high velocities in the core ($R_p \lesssim 1$ Mpc $h^{-1}$) as well as members moving with random motions in the infall regions ($R_p \sim 3 r_v$). 
 
\subsection{Testing the Efficiency of GalWeight on MDPL \& Bolshoi Simulations} \label{sec:simulations}

To further demonstrate and quantify the GalWeight technique  at assigning membership, we again utilize the MDPL2 \& Bolshoi\footnote{https://www.cosmosim.org/cms/simulations/mdpl2/} simulations from the suite of MultiDark simulations. The efficiency of GalWeight can be quantified by calculating two fractions defined as follows. The first is  the completeness $f_c$, which is the fraction of the number of fiducial members identified by GalWeight as members in the projected phase-space relative to the actual number of 3D members projected in the phase-space. The second is the contamination $f_i$, which is the fraction of interlopers  incorrectly assigned to be members, projected in the phase-space (see e.g., \citealp{Wojtak07,Serra13}). Ideally, of course, GalWeight would return fractions of  $f_c = 1$ and $f_i = 0$.

MDPL2 provides us with 1500 simulated clusters with masses ranging from $0.73\times10^{14} h^{-1} M_{\odot}$ to $37.4\times10^{14} h^{-1} M_{\odot}$ to which we can apply GalWeight. We calculate the fractions $f_c$ and $f_i$ at three radii -- $r_v, 2r_v$ and $3r_v$. As shown in Table~\ref{tab:frac}, the mean values of $f_c$ and $f_i$ within $r_v$ are $0.993$ and $0.112$ respectively for the 1500 clusters overall. Also, the fraction $f_c$ decreases from $0.993$ at $r_v$ to $0.981$ at $3r_v$.

For Bolshoi, we have about 500 clusters with masses greater than $0.70\times10^{14} h^{-1} M_{\odot}$. In order to increase the cluster sample of Bolshoi to 1500 clusters, we randomly select different line-of-sights  or ordinations for each distinct halo in additional to the original line-of-sight along the z-direction (see column 3 in Table \ref{tab:frac} for Bolshoi). Then, we apply GalWeight to each cluster. The mass range of the sample is $0.70\times10^{14} h^{-1} M_{\odot}$ to $10.92\times10^{14} h^{-1} M_{\odot}$ as shown in Table~\ref{tab:frac}. The mean values of $f_c$ and $f_i$ within $r_v$ are $0.995$ and $0.126$ respectively for the 1500 clusters overall. Also, the fraction $f_c$ decreases from $0.995$ at $r_v$ to $0.971$ at $3r_v$.

\begin{figure*} \hspace*{-0.0cm}
\includegraphics[width=32.45cm] {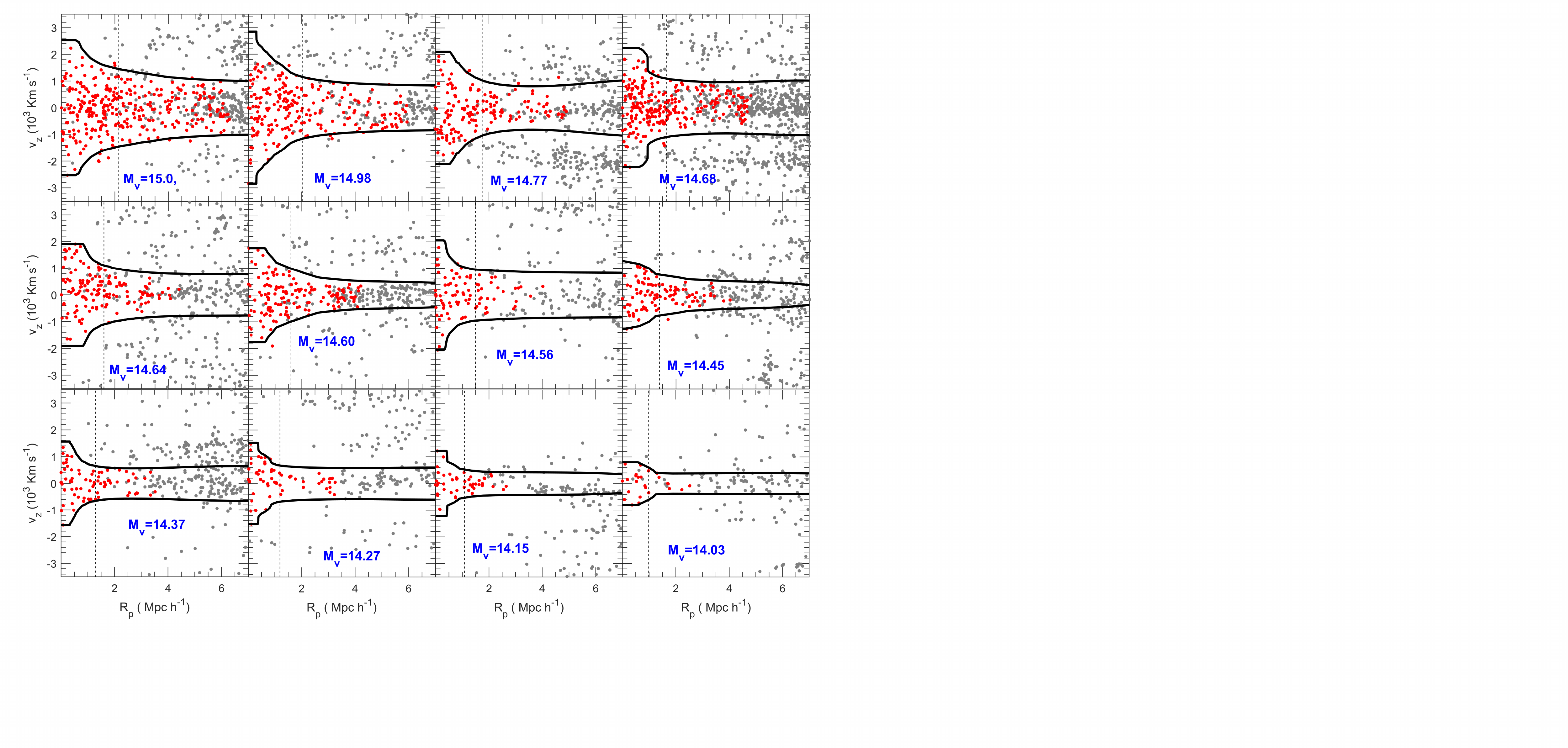} \vspace{-3.05cm}
\caption{Application of the GalWeight technique (solid black lines) to twelve simulated clusters selected from the MDPL simulation (\S~\ref{sec:sims}). Red points show fiducial members within $3r_v$. The virial mass ($\log \mbox{M}_v$ $h^{-1}$ $\mbox{M}_\odot$) and number of members within $r_v$ is shown for each cluster. Clearly, GalWeight does well in effectively identifying members with high accuracy in both the virialized and infall regions for structures ranging in  mass from rich clusters to poor groups. 
}
\label{fig:mocksample}
\end{figure*}

\begin{figure*}\hspace*{-0.2cm}
\includegraphics[width=26cm]{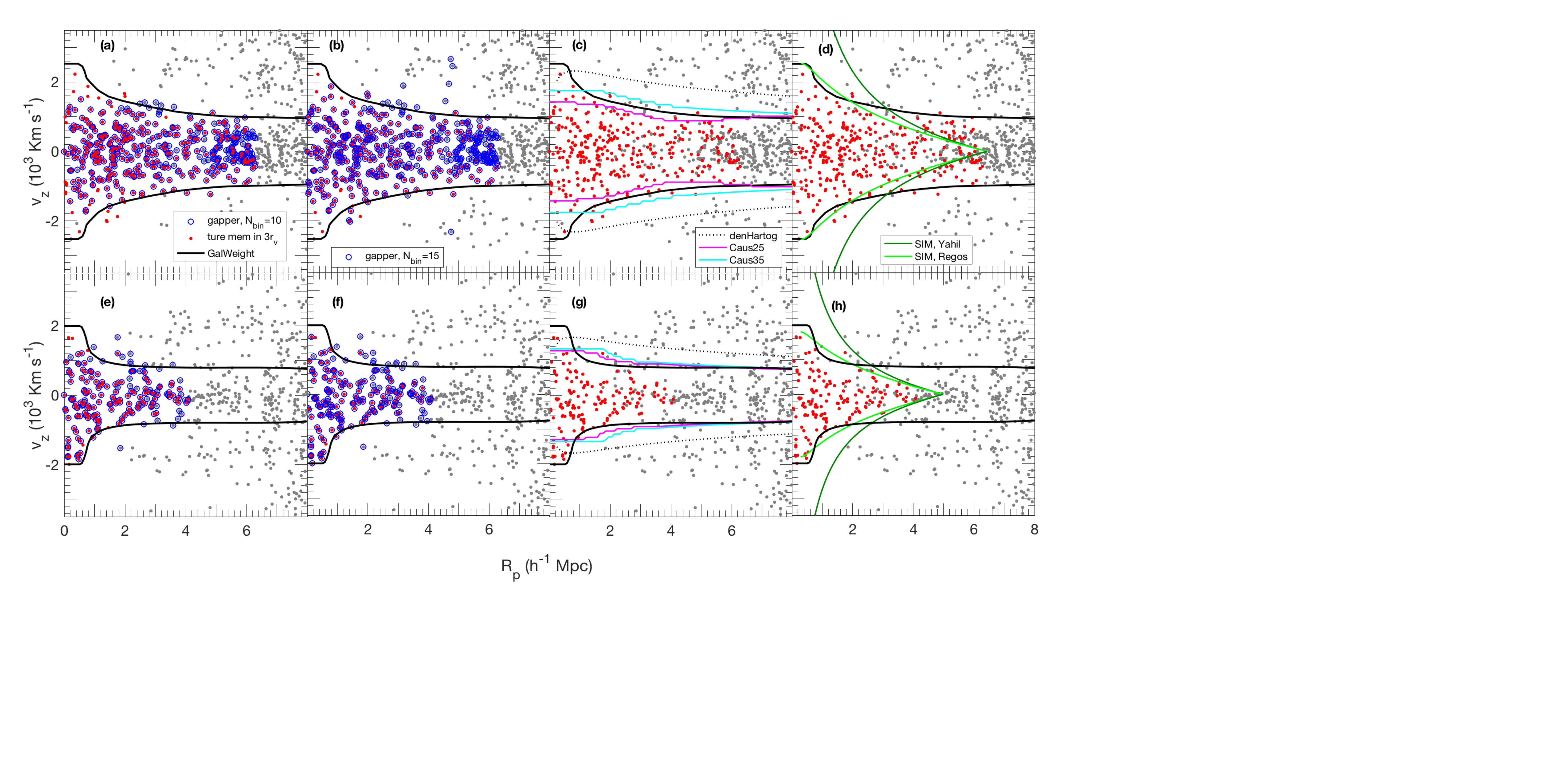} \vspace{-3.75cm}
	\caption{Example of four well-known membership techniques applied to two simulated clusters with mass of  $10.92\times10^{14}$ $h^{-1}$ $\mbox{M}_\odot$ (top panles) \& $4.24\times10^{14}$ $h^{-1}$ $\mbox{M}_\odot$ (bottom) from the Bolshoi simulation (\S~\ref{sec:sims}). In each panel, the red points represent fiducial cluster members within $3r_v$, and the solid black lines show the demarcation contour enclosing cluster members, identified by applying our new technique (GalWeight). The open blue circles in panels (a, b, e \& f) show members identified by the shifting gapper technique using $N_{bin}=10$ and $N_{bin}=15$, respectively. Panel (c \& g) shows the caustic technique employing rescale parameters of q=25 (cyan lines), and q=35 (pink lines) and also the Den Hartog technique (dotted black lines). The Yahil SIM (dark green lines) and Reg\H{o}s SIM (light green lines) techniques are presented in panel (d \& h). GalWeight recovers fiducial members with high accuracy, improving upon the shifting gapper and den Hartog techniques simultaneously at small and large projected radii, the caustic techniques at small projected radius and the SIM technique at large projected radius ($\sim 3r_v$)}.
	\label{fig:Methods1}
\end{figure*}

The main reason that some interlopers are assigned as members ($f_i = 0.113$ for MDPL2 and $f_i = 0.226$ for Bolshoi, as maximal value) is because of the triple-value problem \citep{Tonry81}. That is, there are some foreground and background interlopers that appear to be part of the cluster body because of the distortion of phase-space. The effect of the triple-value problem is apparent in Figure~\ref{fig:mocksample} (discussed below), where most of the interlopers assigned as members are embedded in the cluster body. We defer a discussion of how GalWeight may be adapted to overcome this problem to a future work.

In order to demonstrate the ability of GalWeight to assign membership in the case of both poor and massive clusters we divide the 1500 clusters (for each simulation) into four mass bins as shown in Table \ref{tab:frac}. The fraction $f_c$ varies from 0.998 (0.996) for the poor clusters of mean mass $1.44\times10^{14}~ (1.13\times10^{14})~ h^{-1} M_{\odot}$ to 0.988 (0.997) for the massive clusters of mean mass $11.34\times10^{14}~ (9.68\times10^{14})~ h^{-1} M_{\odot}$ at $r_v$ for MDPL2 (Bolshoi). We conclude that GalWeight can be applied effectively to a range of clusters masses with high efficiency.

Figure~\ref{fig:mocksample} shows examples of GalWeight being applied to twelve simulated Bolshoi clusters (solid black lines), where red and gray points show fiducial members and interlopers, respectively, within $3r_v$. The  twelve clusters shown in Figure~\ref{fig:mocksample} are ranked by virial mass, with the most massive cluster ($10.92\times10^{14}$ $h^{-1}$ $M_{\odot}$) shown in the upper left corner and the least massive one ($1.06\times10^{14}$ $h^{-1}$ $M_{\odot}$) shown in the lower right corner. The figure demonstrates that GalWeight can effectively recover cluster membership for rich massive galaxy clusters as well as small or poor groups of galaxies with the same efficiency.

In summary, applying GalWeight to the suite of MDPL2 and Bolshoi simulations demonstrates that GalWeight can successfully recover cluster membership with high efficiency. It also further demonstrates that it can simultaneously identify members in both the virial and infall regions with taking into account the FOG effect and the random motion of galaxies in the infall region. Furthermore, it can be applied to both rich galaxy clusters and poor groups of galaxies with the same efficiency (see Table \ref{tab:frac}).
\section{A comparison of membership techniques} \label{sec:comp}

In this section, we perform a general comparison between GalWeight and four other well-known techniques ({\bf{shifting gapper, caustic, den Hartog technique, and SIM}}) without doing any quantitative comparison. We defer testing the efficiency of different membership techniques to recover the 3D true members of clusters and the influence of the determining their dynamical masses to a future work (see e.g., \citealp{Wojtak07}).

We begin by showing how each technique fares when it is applied in turn to two simulated clusters with mass of  $10.92\times10^{14}$ $h^{-1}$ $\mbox{M}_\odot$ \& $4.24\times10^{14}$ $h^{-1}$ $\mbox{M}_\odot$ from the  Bolshoi simulation, shown in Figure~  \ref{fig:Methods1}. Making the assumption that the cluster is spherical, fiducial members are assumed to lie within three virial radii, $3r_v$, and are shown as 2D members in the phase-space (red points) in each panel of  Figure~\ref{fig:Methods1}. 
We select this radius ($3r_v$) in order to examine the ability of each technique to recover true members not only within the virial radius but also in the infall region i.e., the region of  a cluster that extends from the viral radius $r_v$ to the turnaround radius $r_t$, where  $r_t \sim ~ 2-4 ~ r_v$. Shown in each panel by the solid black line is the optimal choice of demarcation contour separating members and field galaxies identified by our GalWeight technique. For reasons of space we do not describe each of the four techniques (shifting gapper, den Hartog, caustic and SIM) in detail here. However, we summarize them below and refer the reader to the references for more information.

The {\bf{shifting gapper technique}}  \citep{Fadda96}  works by first placing galaxies into bins according to their projected radial distance from the cluster center. The user has the freedom to choose the number of galaxies per bin which they believe is best-suited to each application of the technique. Commonly chosen  values are $N_{bin}=10$ or 15. For each bin, the galaxies are sorted according to their velocities, then any galaxy separated by more than a fixed value (e.g., $1\sigma$ of the sample or 500-1000 km s$^{-1}$) from the previous one is considered an interloper and removed. \citet{Fadda96} used a gap of 1000 km s$^{-1}$ and a bin of 0.4 $h^{-1}$ Mpc or larger, in order to have at least 15 galaxies. The open blue circles in panels {\bf (a, e) \& (b, f)} of Figure~\ref{fig:Methods1} represent the members identified by this technique, where the number of galaxies utilized per bin was $N_{bin}=10$ and $N_{bin}=15$, respectively. The gray points symbolize interlopers. Clearly, membership identification depends heavily upon the choice of $N_{bin}$, as there are many differences between the galaxies identified as members in panels {\bf (a, e) \& (b, f)}. Additionally, in both cases, some true members of the two cluster are missed, especially at small projected radius. Furthermore, the shifting gapper technique depends on the choice of the velocity gap used to remove interlopers in each bin. A choice of a high-velocity gap results in the identification of large fraction of interlopers as cluster members, while the choice of a low-velocity gap results in missing true cluster members \citep{Aguerri07}. 

The application of the {\bf{caustic technique}} (e.g., \citealp{Alpaslan12,Serra13}) is shown in panels {\bf (c \& g)} of Figure~\ref{fig:Methods1} for two rescale parameters, q = 25 (cyan lines) and q = 35 (pink lines). Although this technique is quite successful when applied to the cluster outskirts, it misses some of the true members located within the core, which are the most important galaxies affecting the dynamics of the clusters. They are missed because the caustic technique does not take into account the effects of the FOG distortion. Also, the caustic technique cannot be applied to small galaxy groups. Furthermore, applying the caustic technique is rather subjective and relies upon how the caustics can be inferred from the data \citep{Reisenegger00, Pearson14}. Nonetheless, it is still a powerful technique for estimating cluster masses.

The application of the {\bf{den Hartog technique}} \citep{denHartog96} is also shown by the dotted black lines in Figure~\ref{fig:Methods1} panels {\bf (c \& g)}. This technique estimates the escape velocity as a function of distance from the cluster center by calculating the virial mass profile (see \S \ref{sec:sdss}), $v_{esc} (R) = \sqrt{\frac{2G M_{vir}(R)}{R}}$, where G is the gravitational constant,. The figure demonstrates that this technique is very biased towards including many far interlopers. In addition, its application relies on assumptions of hydrostatic equilibrium and spherical symmetry.

Panels {\bf (d \& h)} in Figure~\ref{fig:Methods1} show the application of two {\bf{spherical infall models} (SIMs).} The Yahil \citep{Yahil85} and Reg\H{o}s models \citep{Regos89} are shown by dark green and light green lines respectively. Note that, one needs to determine the mass density profile and the background mass density in order to apply the SIM technique and determine the infall velocity profile (e.g., \citealp{vanHaarlem93}). We determine the mass density profile for the simulated cluster from the NFW model (\citealp{NFW96} \& 1997, Equations (\ref{eq:NFW1} \&\ref{eq:NFW11}), knowing its concentration $c$, virial radius $r_v$, and scale radius $r_s = r_v/C$. Also, the background mass density is given by $\rho_{bg} = \Omega_m ~ \rho_c$.

As shown in Figure~\ref{fig:Methods1} (d \& h), SIMs have difficulty identifying true members in the infall region in projected phase-space. This is due to the fact that the effect of random motion of galaxies in the infall regions \citep{vanHaarlem93,Diaferio99} causes some members in the cluster outskirts to be missed. A recent investigation by our own team \citep{Abdullah13} has shown that SIMs can successfully be applied to sliced phase-space by taking into account some kinds of distortions such as the transverse motion of galaxies with respect to the observer and/or rotational motions of galaxies inside the cluster. However, this is out of the scope of the current paper.
\section{Observations - Application to a Sample of 12 Abell Clusters}\label{sec:sdss}

In this section we apply GalWeight to a sample of twelve Abell galaxy clusters, with galaxy coordinates and redshifts taken from SDSS-DR12\footnote{https://http://www.sdss.org/dr12} (hereafter, SDSS-DR12 \citealp{Alam15}). In order to demonstrate the technique for both massive and poor clusters, we selected clusters with Abell richness parameter ranging from 0 to 3 \citep{Abell89}. We deliberately selected some clusters which were almost isolated and others which had clumps or groups of galaxies nearby in order to demonstrate how the technique performs under these different scenarios. We apply the GalWeight technique only to this pilot sample of twelve clusters in this paper, deferring application to the entire SDSS-DR13 sample of $\sim 800$ clusters to a later paper.

The data sample is collected as follows. The NASA/IPAC Extragalactic Database (NED)\footnote{https://ned.ipac.caltech.edu} provides us with a first approximation of the angular coordinates and redshift of the center of our cluster sample ($\alpha_c$, $\delta_c$, $z_c$). We then download the coordinates and redshifts (right ascension $\alpha$, declination $\delta$, and spectroscopic redshift $z$) for objects classified as galaxies near the center of each cluster from SDSS-DR12 \citep{Alam15}. The next step is to apply the binary tree algorithm (e.g., \citealp{Serra11}) to accurately determine the cluster center ($\alpha_{c}$, $\delta_{c}, z_c$) and create a line-of-sight velocity ($v_z$) versus projected radius ($R_p$) phase-space diagram. $R_p$ is the projected radius from the cluster center and $v_z$ is the line-of-sight velocity of a galaxy in the cluster frame, calculated as $ v_z = c( z - z_c)/(1+z_c)$, where $z$ is the observed spectroscopic redshift of the galaxy and $z_c$ is the cluster redshift. The term $(1+z_c)$ is a correction due to the global Hubble expansion \citep{Danese80} and c is the speed of light. 

We then apply GalWeight to the twelve Abell clusters as described in detail in \S~\ref{sec:Tech} in order to get the optimal contour line. The final step is to determine the virial radius, $r_v$, at which $\rho = 200 \rho_c$ and the turnaround radius, $r_t$, at which $\rho = 5.55 \rho_c$  (e.g., \citealp{Nagamine2003,Busha05,Dunner06}) from all galaxies located inside optimal contour line of a cluster.

In order to calculate these two radii we should first determine the cluster mass profile. The cluster mass can be estimated from the virial mass estimator and NFW mass profile \citep{NFW96,NFW97} as follows.

The viral mass estimator is given by

\begin{equation} \label{eq:vir16}
M(<r)=\frac{3\pi N \sum_{i}v_{z, i} (<r)^2}{2G\sum_{i\neq j}\frac{1}{R_{ij}}} 
\end{equation}

\noindent where $v_{z,i}$ is the galaxy line-of-sight velocity and $R_{ij}$ is the projected distance between two galaxies (e.g., \citealp{Limber60,Binney87,Rines03}). 

If a system extends beyond the virial radius, Equation~(\ref{eq:vir16}) will overestimate the mass due to external pressure from matter outside the virialized region \citep{The86,Carlberg97,Girardi98}. The corrected virial mass can be determined using the following expression:

\begin{equation} \label{eq:vir17}
M_{v}(<r)=M(<r)[1-S(r)], \end{equation}

\noindent where $S(r)$ is a term introduced to correct for surface pressure. For an NFW density profile and for isotropic orbits (i.e. the projected, $\sigma_v$, and angular, $\sigma_\theta$, velocity dispersion components of a galaxy in the cluster frame are the same, or equivalently the anisotropy parameter $\beta = 1- \frac{\sigma_\theta^2}{\sigma_r^2} = 0$), $S(r)$ can be calculated by

\begin{equation}\label{eq:vir_25}
S(r)=\left(\frac{x}{1+x}\right)^2\left[\ln(1+x)-\frac{x}{1+x}\right]^{-1}\left[\frac{\sigma_v(r)}{\sigma(<r)}\right]^2,
\end{equation}

\begin{figure*} \hspace*{0.0cm}
\includegraphics[width=32cm]{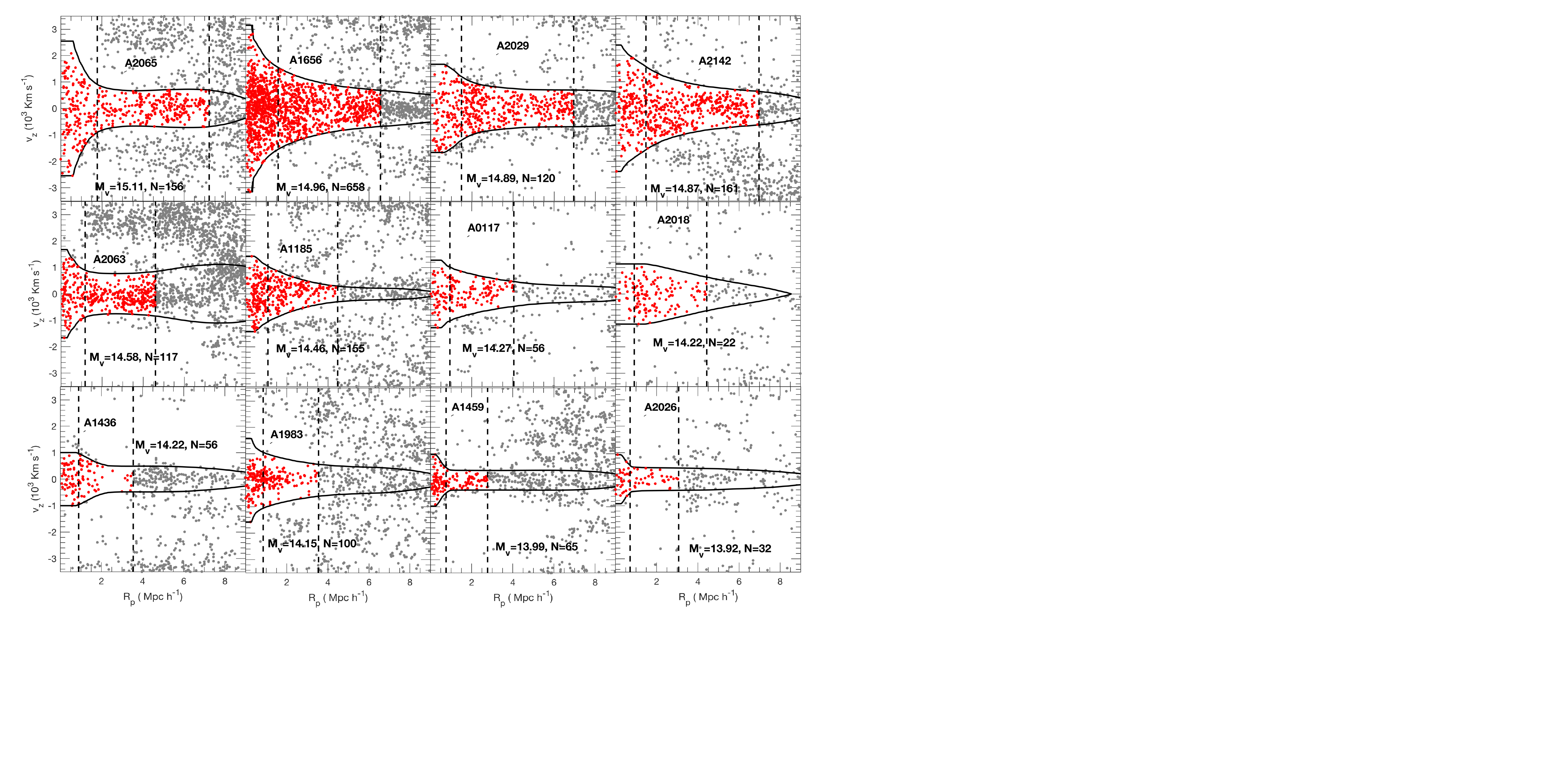} \vspace{-4cm}
\caption{Application of the GalWeight technique to twelve Abell clusters from SDSS-DR12 (see also Table~\ref{tab:parameters}). The solid black lines shows the optimal contour line and the two dashed vertical lines show the virial and turnaround radii respectively. The red points show galaxies identified as clusters members - those enclosed by optimal contour line and $r_t$. Also shown in each panel is the cluster virial mass ($\log \mbox{M}_v$ $h^{-1}$ $\mbox{M}_\odot$) and number of galaxies within $r_v$.
}
\label{fig:All}
\end{figure*}

\noindent where $x=r/r_s$, $r_s$ is the scale radius, $\sigma(<r)$ is the integrated three-dimensional velocity dispersion within $r$, and $\sigma_v(r)$ is a projected velocity dispersion (e.g., \citealp{Koranyi2000,Abdullah11}).

The mass density within a sphere of radius $r$ introduced by NFW is given by

  \begin{equation} \label{eq:NFW1}
  \rho(r)=\frac{\rho_s}{x\left(1+x\right)^2}, \end{equation}

\noindent and its corresponding mass is given by

   \begin{equation} \label{eq:NFW11}
   M(<r)=\frac{M_s}{\ln(2)-(1/2)}\left[\ln(1+x)-\frac{x}{1+x}\right],
   \end{equation}

\noindent where $M_s=4\pi\rho_s r^3_s [\ln(2)-(1/2)]$ is the mass within $r_s$, $\rho_s = \delta_s \rho_c$ is the characteristic density within $r_s$ and $\delta_s = (\Delta_v/3) c^3 \left[\ln(1+c) - \frac{c}{1+c}\right]^{-1}$, and the concentration $c = r_v/r_s$  (e.g.,  \citealp{NFW97,Rines03,Mamon13}).

\begin{table*} \centering
\caption{Dynamical parameters derived for the sample of twelve Abell galaxy clusters}
\label{tab:parameters}
\scriptsize
\begin{tabular}{cc cccccc c cccccc c ccc}\hline
cluster &  $z_c$ &\multicolumn{6}{c}{virial mass estimator} &&  \multicolumn{6}{c}{NFW mass estimator} &&  \multicolumn{3}{c}{NFW parameters}\\
\cline{3-8}  \cline{10-15} \cline{17-19}\\
&&$r_{500}$&$M_{500}$&$r_{200}$ &  $M_{200}$&$r_{100}$&$M_{100}$   &&   $r_{500}$&$M_{500}$&$r_{200}$ &  $M_{200}$&$r_{100}$&$M_{100}$&&$r_{s}$&$M_{s}$ & c \\
(1)&(2) &(3)&(4)&(5)&(6)&(7)&(8)  &&  (9)&(10)&(11)&(12)&(13)&(14) && (15)&(16)&(17)\\ 
\hline
A2065 & 0.073 & 1.27 &11.95 & 1.78 &12.97& 2.29&13.86 && 1.20 &10.11 & 1.78 &13.01 & 2.36 &15.30 && 0.22 & 1.90 & 8.16\\
A1656 & 0.023 & 1.07 & 7.14 & 1.58 & 9.06& 2.07&10.30 && 1.06 & 6.78 & 1.58 & 9.09 & 2.11 &10.96 && 0.26 & 1.61 & 6.01\\
A2029 & 0.078 & 0.99 & 5.68 & 1.49 & 7.65& 1.97& 8.84 && 0.94 & 4.73 & 1.49 & 7.67 & 2.07 &10.31 && 0.61 & 2.80 & 2.46\\
A2142 & 0.090 & 0.97 & 5.24 & 1.47 & 7.27& 2.03& 9.64 && 0.91 & 4.35 & 1.47 & 7.32 & 2.05 &10.02 && 0.67 & 2.99 & 2.19\\
A2063 & 0.035 & 0.80 & 2.91 & 1.17 & 3.73& 1.54& 4.23 && 0.81 & 3.10 & 1.18 & 3.76 & 1.55 & 4.27 && 0.08 & 0.40 & 14.40\\
A1185 & 0.033 & 0.75 & 2.38 & 1.08 & 2.89& 1.42& 3.29 && 0.72 & 2.20 & 1.08 & 2.91 & 1.44 & 3.48 && 0.17 & 0.49 & 6.53\\
A0117 & 0.055 & 0.65 & 1.55 & 0.93 & 1.86& 1.26& 2.31 && 0.61 & 1.29 & 0.93 & 1.88 & 1.27 & 2.37 && 0.24 & 0.46 & 3.87\\
A2018 & 0.088 & 0.55 & 0.92 & 0.90 & 1.67& 1.35& 2.84 && 0.55 & 0.95 & 0.90 & 1.67 & 1.27 & 2.36 && 0.48 & 0.82 & 1.83\\
A1436 & 0.065 & 0.49 & 0.68 & 0.89 & 1.64& 1.25& 2.27 && 0.61 & 1.29 & 0.89 & 1.64 & 1.18 & 1.92 && 0.10 & 0.23 & 8.68\\
A1983 & 0.045 & 0.58 & 1.09 & 0.85 & 1.39& 1.09& 1.51 && 0.57 & 1.03 & 0.85 & 1.41 & 1.14 & 1.71 && 0.16 & 0.27 & 5.37\\
A1459 & 0.020 & 0.50 & 0.73 & 0.72 & 0.85& 0.92& 0.89 && 0.50 & 0.73 & 0.72 & 0.87 & 0.95 & 0.97 && 0.04 & 0.08 & 18.4\\
A2026 & 0.091 & 0.46 & 0.54 & 0.71 & 0.82& 0.91& 0.88 && 0.47 & 0.59 & 0.71 & 0.83 & 0.96 & 1.03 && 0.16 & 0.19 & 4.32\\
\hline
\end{tabular}
\begin{tablenotes}
\item
\noindent Radii and their masses are calculated by virial and NFW mass estimators at overdensities of $\Delta = 500$, $200$ and 100 $\rho_c$. The radius and mass are in units of $h^{-1} \mbox{Mpc}$ and  $10^{14}$ $h^{-1} \mbox{M}_\odot$.\\
Columns: (1) cluster name; (2) cluster redshift; (3-4), (5-6) \& (7-8) are radii and their corresponding masses  calculated bythe  virial mass estimator at overdensities of $\Delta= 500, 200$ and 100, respectively. (9-10), (11-12) \& (13-14) are radii and their corresponding masses calculated by an NFW model at overdensities of $\Delta= 500, 200$ and 100, respectively. (15-17) are scale radius, its corresponding scale mass and concentration of NFW parameters.
\end{tablenotes}
\end{table*} 

The projected number of galaxies within a cylinder of radius R is given by integrating the NFW profile (Equation~(\ref{eq:NFW1})) along the line of sight (e.g., \citealp{Bartelmann96,Zenteno16})

   \begin{equation} \label{eq:NFW2}
   N(<R)=\frac{N_s}{\ln(2)-(1/2)} g(x),
   \end{equation}

\noindent where $N_s$ is the number of galaxies within $r_s$ that has the same formula as $M_s$, and $g(x)$ is given by (e.g., \citealp{Golse02,Mamon10})

   \begin{equation}
   g(x) = \begin{cases} \ln(x/2) + \frac{\cosh^{-1} (1/x)}{\sqrt{1-x^2}} \ \ \ \mbox{if} \  x \ < \ 1\\ 
	                      1-\ln(2)  \ \ \ \ \ \ \ \ \ \ \ \ \ \ \ \ \ \  \  \mbox{if} \  x \ = \ 1 \\ 
	                      \ln(x/2) + \frac{\cos^{-1} (1/x)}{\sqrt{x^2-1}}  \ \  \ \ \ \mbox{if} \  x \ > \ 1
   \end{cases} \end{equation}

Thus, we can fit $r_s$ for each cluster to get $S(r)$ from Equation~\ref{eq:vir_25} and calculate the corrected mass profile $M_v(r)$ from Equation~\ref{eq:vir17}. Also, the NFW mass profile is calculated from Equation~\ref{eq:NFW11}. Then, $r_v$, at which $\Delta = 200 \rho_c$, can be calculated from the viral or NFW mass profiles. While $r_t$, at which $\Delta = 5.55 \rho_c$, can be determined from NFW mass profile only. We cannot determine $r_t$ from the virial mass profile because the assumption of hydrostatic equilibrium is invalid.

Finally, after we calculate $r_v$ and $r_t$ (from NFW mass profile) the cluster membership can be defined as all galaxies enclosed by the optimal contour line and within $r_t$, as shown by the red points in Figure~\ref{fig:All}. It is worth noting once again that GalWeight is effective at taking into account the effects of the FOG distortion in the innermost regions and the random motion of galaxies in the cluster infall region. Moreover, GalWeight is not affected by the presence of substructures or nearby clusters or groups as demonstrated, for example, for A2063 \& A2065. Furthermore, GalWeight can be applied both to rich clusters such as A2065 \& A1656 and to poor galaxy groups such as A1459 \& A2026.

In order to compare our results with the literature, we calculate the radii and their corresponding masses at three overdensities, $\Delta_{500} = 500 \rho_c$, $\Delta_{200} = 200 \rho_c$ and $\Delta_{100} = 100 \rho_c$ as shown in Table \ref{tab:parameters}. The sample is displayed in order of decreasing NFW $M_{200}$ mass. A complete list of NFW parameters is also provided in Table \ref{tab:parameters}. 

In Table \ref{tab:comp} we list ratios of radii and masses for each of the twelve Abell clusters using our GalWeight-determined method (assuming an NFW profile) divided by previously-published values, $(r_{NFW}/r_{ref})$ and $(M_{NFW}/M_{ref})$ respectively, at overdensities of $\Delta = 500$, $200$ and $100 \rho_c$. Column 8 of Table~\ref{tab:comp} also lists the ratio of GalWeight-determined masses relative to those estimated from the caustic technique \citep{Rines16}, ($M_{NFW}/M_{caus})_{200}$, at $\Delta= 200 \rho_c$. Table~\ref{tab:comp} clearly shows that the radii and masses estimated  for a cluster are strongly dependent on the technique used to assign membership and remove interlopers (see \citealp{Wojtak07}). The ratio $(r_{NFW}/r_{ref})$ ranges between 0.63 and 1.55, while the ratio $(M_{NFW}/M_{ref})$ ranges between 0.58 and 2.18. 

\begin{table*} \centering
\caption{GalWeight-determined ratios of radii and mass for each of the twelve Abell clusters compared to previously-published values}
\label{tab:comp}
\scriptsize 
\begin{tabular}{cccccccccccc}\hline
cluster &  \multicolumn{3}{c}{$(r_{NFW}/r_{ref})$} &&&  \multicolumn{3}{c}{$(M_{NFW}/M_{ref})$} &&&  \multicolumn{1}{c}{($M_{NFW}/M_{caus})_{200}$}\\
\cline{2-4}  \cline{7-9} \\
(1)& (2) &(3)&(4)&&&(5)&(6)&(7)&&&(8)\\ 
&\multicolumn{1}{c}{$\Delta_{500}$}&\multicolumn{1}{c}{$\Delta_{200}$}& \multicolumn{1}{c}{$\Delta_{100}$}&&&\multicolumn{1}{c}{$\Delta_{500}$}&\multicolumn{1}{c}{$\Delta_{200}$}&\multicolumn{1}{c}{$\Delta_{100}$}&&&\multicolumn{1}{c}{\citet{Rines16}}\\
\hline
A2065 & $1.70^{5}$, $1.15^{8}$  & $1.04^{8}$, $1.11^{11}$  			  &  ---           &&& $1.50^{8}$             & $1.12^{8}$, $1.30^{11}$  &  ---        &&&  3.84 \\
A1656 & $1.16^{5}$              & $0.89^{4}$, $1.05^{9}$   		 	  & $1.04^{6}$     &&& $1.09^{5}$             & $0.72^{4}$, $1.16^{9,+}$   & $1.12^{6}$  &&&  2.02 \\
A2029 & $1.05^{5}$, $0.93^{8}$  & $0.93^{8}$, $0.89^{11}$  			  &  ---           &&& $0.92^{5}$, $0.80^{8}$ & $0.82^{8}$, $0.88^{11}$  &  ---        &&&  1.42 \\
A2142 & $0.91^{8}$, $1.30^{10,+}$ & $1.30^{10,+}$, $0.94^{11}$ 			  & $0.96^{12}$    &&& $0.66^{8}$             & $1.80^{10,+}$, $0.75^{11}$ & $0.86^{12}$ &&&  2.27 \\
A2063 & $1.30^{8}$              & $1.18^{8} $              			  & $1.05^{12}$    &&& $2.18^{8}$             & $1.65^{8}$               & $1.03^{12}$ &&&  1.40 \\
A1185 &  ---                    & $1.01^{3,\bullet} $              &  ---           &&& ---                    & $2.77^{3,\bullet}$               &  ---        &&&  1.37 \\
A0117 &  ---                    & $0.83^{1,\star}$, $1.05^{2} $  	  &  ---           &&& ---                    & $0.58^{1,\star}$               &  ---        &&&  ---  \\
A2018 &  ---                    & $0.82^{2} $, $1.18^{13}$ 			  &  ---           &&& ---                    &  ---                     &  ---        &&&  0.94 \\
A1983 & $0.85^{7}$              & $0.90^{3,\bullet} $, $0.81^{7}$  &  ---           &&& $1.18^{7}$             & $0.99^{3,\bullet}$,$1.03^{7}$   &  ---        &&&  1.64 \\
A1436 & $1.55^{10,+}$             & $0.64^{1,\star}$, $1.24^{10,+}$  	  &  ---           &&& ---                    & $0.64^{1,\star}$, $1.30^{10,+}$  &  ---        &&&  ---  \\
A1459 &  ---                    & $0.94^{1,\star} $              	  &  ---           &&& ---                    & $0.84^{1,\star}$               &  ---        &&&  ---  \\
A2026 &  ---                    & $0.63^{2} $              			  &  ---           &&& ---                    &  ---                     &  ---        &&&  ---  \\
\hline
\end{tabular}

\begin{tablenotes}
\item
Columns: (1) cluster name; (2-4) ratio of GalWeight radii to those in the literature at overdensities of $\Delta = 500$, $200$ and $100$ $\rho_c$ respectively, assuming an NFW model. (5-7) ratio of GalWeight masses  to those in the literature at overdensities of $\Delta = 500$, $200$ and $100$ $\rho_c$ respectively, assuming an NFW model. (8) ratio of GalWeight masses assuming an NFW model to those calculated from the caustic technique in \citet{Rines16} at $\Delta= 200$ $\rho_c$ .\\
1=\citet{Abdullah11}, 2=\citet{Aguerri07}, 3=\citet{Girardi02}, 4=\citet{Kubo07}, 5=\citet{Lagana11}, 6=\cite{Lokas03}, 7=\citet{Pointecouteau05}, 8=\citet{Reiprich02}, 9=\citet{Rines03}, 10=\citet{Rines06}, 11=\citet{sifon15}, 12=\citet{Wojtak10}. $^+$ (caustic technique), 
$^\bullet$ (shifting gapper),$^\star$ (SIM).
\end{tablenotes}
\vspace{4mm}
\end{table*} 


The cluster masses from the literature tabulated in Table~\ref{tab:comp}  have been calculated in various ways. Below, we explicitly compare 
our values to those obtained from applying the shifting gapper, SIM and caustic methods.

First, comparing to the shifting gapper technique (see ($^\bullet$) in Table~\ref{tab:comp},  \citealp{Girardi02,sifon15}), we find that the ratio $(M_{NFW}/M_{ref})$ is larger than unity in some cases (A2065, A1185) and smaller than unity in others (A2029, A2142). This is because members assigned by this technique, and consequently the mass calculated, depend on the selection criteria of number of galaxies and velocity gap per bin. As discussed before, the choice of a high-velocity gap includes more members and consequently larger mass and vice versa.

Second, comparing to the SIM method (see ($^\star$) in Table~\ref{tab:comp}, \citealp{Abdullah11}) we note that the mass ratio $(M_{NFW}/M_{ref})$ is less than unity for the three clusters A0117, A1436 and A1459. This is because SIM includes more galaxy members inside the virial region even though they are very far from the cluster body. This is due to the assumption of conservation of mass that influences on the validity of SIM in the innermost region (see Figure 6 in \citealp{Abdullah11}).

\begin{figure*}\hspace*{0.0cm}
	\includegraphics[width=26.5cm]{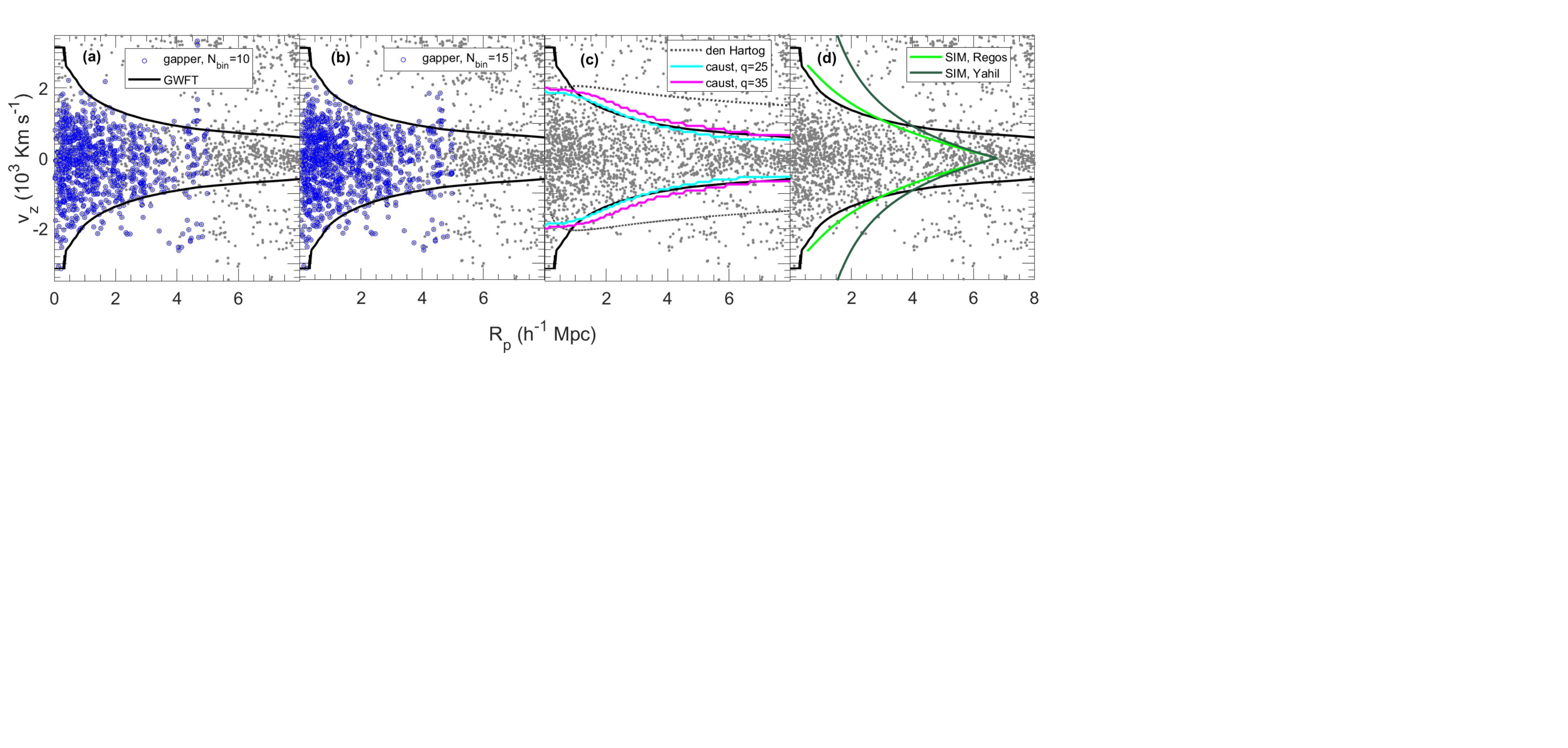} \vspace{-7.0cm}
	\caption{Example of four well-known membership techniques applied to the 
	cluster. The blue open symbols and solid lines are as in Figure~\ref{fig:Methods1}.  Clearly GalWeight (solid black lines) appears to identify cluster members well both in the virialized and infall regions of phase-space.}
	\label{fig:Methods2}
\end{figure*}

Third, comparing to the caustic technique (see ($^+$) in Table \ref{tab:comp}, \citealp{Rines03,Rines06,Rines16}) we specifically calculate the ratio $(M_{NFW}/M_{caus})_{200}$ as listed in Table~\ref{tab:comp}, column 8. It demonstrates that this ratio is larger than unity for 7 clusters with the highest ratio is for A2065, for which the estimated mass from NFW is four times that expected from the caustic technique. As described above, the main reason for this discrepancy is that the caustic technique does not take into consideration the effect of FOG. Thus, it misses more members inside the virial region and consequently expects lower masses.

We compare again GalWeight with the four well-known techniques (shifting gapper, caustic, den Hartog, and SIM) for the Coma cluster as shown in Figure~\ref{fig:Methods2}. The Figure (see also Figure~\ref{fig:Methods1}) demonstrates that the GalWeight performs very favorably against established methods, taking into account as it does the effects of the FOG distortion at small projected radius well as the random motion of galaxies in the infall region. In order to apply SIM to the Coma cluster the spatial number density profile is calculated from the NFW model (\citealp{NFW96,NFW97}). Also, we assume that the background number density $\rho_{bg} = 0.0106$ $h^3$ $\mbox{Mpc}^{-3}$ which is calculated using the parameters of Schechter luminosity function ($\phi^\ast = 0.0149$ $h^3$ $\mbox{Mpc}^{-3}$, $\mbox{M}^\ast -5\log{h} = -20.44$ and $\alpha = -1.05$ for $r$ magnitude, \citealp{Blanton03}).

Because of the presence of interlopers, estimates of cluster mass tend to be biased too high and estimates of cluster concentration tend to be biased too low. Our work suggests that applying GalWeight rather than another technique to determine cluster membership before applying a dynamical mass estimator (virial theorem, NFW model etc.), likely results in a more accurate estimate of the true cluster mass and concentration. In  a future work we will compare the efficiency of different membership techniques to assign membership and their influence on estimating cluster mass using different  mass estimators.

\section{Discussion and Conclusion} \label {sec:conc}

In this paper we introduced the Galaxy Weighting Function Technique (GalWeight), a powerful new technique for identifying cluster members. specifically designed to simultaneously maximize the number of {\it bona fide} cluster members while minimizing the number of contaminating interlopers.

GalWeight takes into account the causes of different distortions in phase-space diagram and is independent of statistical or selection criteria. 
It can recover membership in both the virial and infall regions with high accuracy and is minimally affected by substructure and/or nearby clusters.

We first demonstrated GalWeight's use by applying it interactively to a simulated cluster of mass $9.37 \times 10^{14}$ $h^{-1}$ M$_{\odot}$ selected from Bolshoi simulation. 
Next, we tested the efficiency of the technique on $\sim 3000$ clusters selected from the MDPL2 and Bolshoi simulations with masses ranging from $0.70\times10^{14} h^{-1} M_{\odot}$ to $37.4\times10^{14} h^{-1} M_{\odot}$. The completeness and interloper fractions for MDPL2 are $f_c = 0.993, 0.992$ and $0.981$ and $f_i = 0.096, 0.098$ and $0.118$, while for Bolshoi  $f_c = 0.995, 0.981$ and $0.971$ and $f_i = 0.126, 0.217$ and $0.226$ within $r_v$, $2r_v$ and $3r_v$, respectively. We then compared its performance to four well-known existing cluster membership techniques (shifting gapper, den Hartog, caustic, SIM). Finally, we applied GalWeight to a sample of twelve Abell clusters of varying richnesses taken from SDSS-DR12. By assuming an NFW model and applying the virial mass estimator we determined the radius and corresponding mass at overdensities of $\Delta_{500}$, $\Delta_{200}$ and $\Delta_{100}$. The virial mass (at $\Delta_{200}$) of the sample ranged from $0.82\times10^{14}$ $h^{-1}$ $M_{\odot}$ to $12.97\times10^{14}$ $h^{-1}$ $M_{\odot}$, demonstrating that GalWeight is effective for poor and massive clusters. In the future we plan to apply GalWeight to a larger SDSS sample of galaxy clusters at low and high redshifts.

We believe that GalWeight has the potential for astrophysical applications far beyond the identification of  cluster members e.g., identifying stellar members of nearby dwarf galaxies, or separating star-forming and quiescent galaxies. We also plan to investigate these applications in a future work. 

\section*{Acknowledgement}
We thank Brian Siana for useful discussions. We also thanks Gary Mamon for his useful comments on the paper. Finally, we thank the reviewer for suggestions which improved this paper.
G.W. acknowledges financial support for this work from NSF grant AST-1517863 and from NASA through programs GO-13306, GO- 13677, GO-13747 \& GO-13845/14327 from the Space Telescope Science Institute, which is operated by AURA, Inc., under NASA contract NAS 5-26555, and grant number 80NSSC17K0019 issued through the Astrophysics Data Analysis Program (ADAP).

\end{document}